# Water saturation in texturally porous carbonate rocks: shock thermodynamics and dampening of the shock.

Juulia-Gabrielle Moreau[1,*], Argo Jõeleht[1], Anna Losiak[2,3], Meng-Hua Zhu[4], Jüri Plado[1]

[1]Department of Geology, University of Tartu, Ravila 14A, 50411 Tartu, Estonia
[2]Institute of Geological Sciences, Polish Academy of Sciences, Podwale 75, PL-50449 Wroclaw, Poland
[3]Lunar and Planetary Institute, Houston, USA
[4]State Kay Laboratory of Lunar and Planetary Sciences, Macau University of Science and Technology, Taipa, Macau, China

**Abstract**

Sedimentary rocks often form the upper layers or the entirety of target rocks in impact events. Thermodynamic properties of sedimentary rocks related to porosity and water saturation affect the process of impact crater formation. The heterogeneous distribution of sedimentary facies can complicate shock-effects development and distribution, especially in numerical modeling. This work focuses on the shock thermodynamic properties of carbonate rocks with differing porosity textures (e.g., isolated pores, interstitial porosity, elongated pores), and water saturation levels. Using mesoscale numerical modeling, we found that water saturation reduces shock temperatures compared to those in dry, porous carbonate rocks. The orientation of elongated pores and porosity lineations influences the shock temperature distribution and rock deformation at angles of 50–90° to the shock front. Additionally, interstitial porosity plays a key role in creating temperature zonations around larger grains, owing to complex shock wave interactions.

* Corresponding author. e-mail address: juulia.moreau@ut.ee / permanent: jmoreauulb@gmail.com, postal address: Ravila 14A, 50411 Tartu, Estonia

# 1. Introduction

Impact processes are a common occurrence in the Solar System and play a crucial role in the formation and evolution of planetary bodies, including planetary accretion (Montmerle et al., 2006). This is witnessed by the numerous impact craters visible on the surfaces of rocky planets and moons, such as the Moon (Robbins, 2018), Mercury (Herrick et al., 2018), asteroids (Marchi et al., 2016), Mars (Cabrol and Grin, 1999), and Earth (Kenkmann, 2021, Osinski et al., 2022). As of 2020, the confirmed count of impact craters on Earth stands at nearly 200 (Gottwald et al., 2020; Kenkmann, 2021), with about half located in North America, Northern Europe, and Russia. About 57% of terrestrial impacts have occurred in sedimentary rocks, with 25% of all such structures in carbonate rocks (limestone and dolostone targets). Sedimentary impact targets are often heterogeneous and porous, consisting of intercalating layers of various lithologies (e.g., Odessa, Texas, Littlefield et al., 2007; Morasko, Poland, Szokaluk et al., 2019; Kaali, Estonia, Plado, 2012, Losiak et al. 2016, 2022). On a smaller scale, carbonate rocks develop diverse pore textures (Lucia, 2004) due to diagenetic processes (e.g., dissolution, cementation, dolomitization, Larsen and Chilingor, 1979), rock composition (e.g., fossiliferous, crystalline, Dunham, 1962), subsurface weathering (Ott et al., 2023), cavity formation (e.g., karst systems, Ford and Williams, 2007), and depositional history (Clyde and William, 2013). The pore textures vary widely, including interparticle and intraparticle porosity, channels, fractures, and fenestral and vuggy pores (Janjuhah et al., 2021; Plummer, 2021). The formation of pore space in carbonate rocks through synsedimentary or diagenetic processes will affect their rheology. This pore space may eventually be saturated with water, which plays a key role in crater formation, potentially leading to features such as lower rims, terracing, and spalling (Hoerth et al., 2013; Takita and Sumita, 2013; Dufresne et al., 2013; Kenkmann et al., 2018). While the extent of water saturation at the time of impact is not always clear, it is well-documented for certain craters with known water levels in the target rock (e.g., Kaali, Veski et al., 2007, Losiak et al., 2016, Rosentau et al., 2021). Furthermore, water saturation has been shown to affect the thermodynamic properties of the rock under shock, influencing relationships such as shock wave velocity, particle velocity, and volume changes under shock pressure (*PdV*) (sandstone, Arlery et al., 2010; tuff, Anderson and Larson, 1977).

Research on the shock thermodynamics of carbonate rocks often involves shock experiments on dry, porous, homogeneous, and well-crystallized specimens (Martinez et al., 1995; Hörz et al., 2019; Baranowski et al., 2022; Kurosawa et al., 2022), or analytical equations of state models for water-saturated limestone (Brown, 1989). However, the role of water saturation, textural diversity of the pore space, or the combination of the two, has never been studied, especially with numerical modeling and shock experiments, or to establish a detailed shock classification of carbonate rocks involving benchmarking between modeling, experiments, and observations (Stöffler et al., 2018).

This study employs numerical simulations to investigate shock thermodynamics resulting from water-saturation and pore space textures of carbonate rocks. Because pore space is essentially observable at the millimeter scale or that meter-scale or kilometer-scale impact modelling cannot replicate textural heterogeneities at the millimeter-scale resolution, a mesoscale modelling approach must be considered (Crawford et al., 2003). The mesoscale modelling is thus used to i) re-evaluating thermodynamic data for water-saturated carbonates, ii) simulating the effects of partial and full water saturation, iii) analyzing the role of water in shock heating, and iv) underline the importance of varied pore configurations, such as isolated and interstitial pores, pore abundance, orientation of elongated pores (e.g., fenestral porosity), and open fractures in dry or water-saturated carbonates. We aim to demonstrate that the heterogeneity of carbonate rock porosity and the presence of fluid-filled pores influence the distribution of shock temperatures and create potential breaking points within shock dynamics. This highlights the need for case-specific studies in sedimentary rock, especially carbonate rock shock metamorphism, providing groundwork toward a reliable way to classify carbonate rocks as a function of their level of shock and water saturation (Stöffler et al., 2018).

## 2. Methods

To simulate the effects of water saturation on porous carbonate rocks, we utilized the iSALE 2-D shock physics code (Dellen version; Wünnemann et al., 2006; Collins et al., 2012). The code builds on the original SALE hydrocode (Amsden et al., 1980) designed for hypervelocity impact simulations in solid materials. SALE has been upgraded to incorporate an elastic-plastic constitutive model, fragmentation models, various equations of state (EoS), and the capability to simulate multiple materials simultaneously (Melosh et al., 1992; Ivanov et al., 1997). More recent updates feature a modified strength model (Collins et al., 2004) and a porosity compaction model (Wünnemann et al., 2006; Collins et al., 2011).

### *2.1. Material properties*

We used the calcite ANEOS material model to simulate carbonate rocks in general, allowing for accurate extraction of shock thermodynamic properties (Lucia, 2004). Literature data were applied to match the strength and thermal properties of dolostone to the calcite ANEOS model to represent more competent carbonate rocks, as outlined in Table S1, including i) cohesion of intact material (Patel and Shah, 2015), ii) limited strength at high pressure (Kurosawa et al., 2022), iii) coefficient of friction for intact material (Demurtas et al., 2021), iv) coefficient of friction for damaged material (Scuderi et al., 2013), and v) heat capacity (Eppelbaum et al., 2014). Unlike some impact models, we did not implement an ε–α compaction model (Wünnemann et al., 2006; Collins et al., 2011) to account for porosity within the material, making the simulated carbonate rock nonporous. Instead, porosity was modeled separately using empty cell clusters (particles). Water-filled pores were simulated using the water ANEOS, while dry pores were represented by empty numerical cells, simulating air.

**Table 1.** Parametrization of numerical models

| #Fig. 1 | model | cell (µm) | width (µm) | thickness (µm)[a] | | | | inclusions | velocity (m/s) | nominal pressure (GPa) | symmetry | CPPR[b] | BMP |
|---|---|---|---|---|---|---|---|---|---|---|---|---|---|
| | | | | FP | BL | SL | BL | | | | | | |
| b | water-saturated c.-rock | 1 | 500 | 750 | 500 | 500 | 500 | 10/20/40% water-filled pores | 1000–8000 | 4–96 | cylindrical[c] | 20±4 | random |
| b | partially saturated c.-rock | 1 | 500 | 750 | 500 | 500 | 500 | 18% water-filled pores + 15% empty pores | 1000–8000 | 4–96 | cylindrical[c] | 20±4 | BL/SL/BL |
| b | dry c.-rock | 1 | 500 | 750 | 500 | 500 | 500 | 17/33% empty pores | 1000–8000 | 4–96 | cylindrical[c] | 20±4 | BL/SL/BL |
| b | pure c.-rock | 1 | 50 | 76 | 50 | 50 | 50 | - | 1000–15,000 | 4–269 | cylindrical | - | - |
| b | water | 1 | 25 | 300 | 30 | 30 | 30 | - | 1000–25,000 | 1–187[d] | cylindrical | - | - |
| c | single pore in c.-rock | 1 | 150 | 230 | 230 | 230 | 460 | dry or saturated pore | 30 | 0.1 | cylindrical | 27 | SL |
| d | oriented pore in c.-rock | 4 | 3200 | 2400 | 2000 | 2280 | 2000 | 0–90° dry or saturated elongated crack[e] | 3000 | 20 | planar | 18×210 | SL |
| e | simplified fenestral porosity | 4 | 2400 | 4800 | 2000 | 2540 | 2000 | 0–90° dry discontinuous crack lineations[e] | 1500–4000 | 8–32 | planar | 10×160 | SL |
| f | open fracture and interstitial porosity in c.-rock[f] | 4 | 1600 | 4800 | 2000 | 3360 3520 3720 | 2000 | 87% c.-rock in saturated or dry pore space | 3000 | 20 | planar | ~8–30 | SL |

(a) FP: flyer plate; BL, buffer layer; SL, sample layer
(b) cells per particle radius or total cells widths (...×...)
(c) replicated in non-cylindrical models (20% water, partially saturated, dry)
(d) in water
(e) angle to the shock wave front
(f) 120, 280, 480 µm thick fractures at sample centre

*2.2. Setup for numerical models*

We employed a mesoscale approach to study the effects of water saturation on porous carbonate rocks, drawing on previous studies (Crawford et al., 2003; Ivanov, 2005; Riedel et al., 2008; Borg and Chhabildas, 2011; Güldemeister et al., 2013; Bland et al., 2014; Davison et al., 2016, 2017; Moreau et al., 2018, 2019; North et al., 2023). The model was configured with 4 planar layers (Fig. 1). The topmost layer represented the impactor/flyer plate, with the impact velocity determining the pressure input across the model. The second layer acted as a buffer to distribute a steady shock wave into the third layer (the sample layer) and delayed the rarefaction wave from the flyer plate. The third layer, representing the sample, had various configurations (Fig. 2b–f). The fourth layer served as a buffer to prevent early shock decay due to the model boundary conditions. All layers were virtually made of the same background material (carbonate rock), including the flyer plate, with the sample layer containing voids that were either empty or filled with water, randomly distributed, or arranged to a predefined bitmap file (iSALE modification; Moreau et al., 2019) (Fig. 1b–f).

The resolution for each configuration (Table 1) is provided in cells per (void) particle radius (CPPR), along with the corresponding cell size. The CPPR values (~10–30 CPPR) used in this study were adequate (Moreau et al., 2017; showing 1–2% error on peak shock pressures within particles, Moreau et al., 2019) to capture localized pressure increase around and within the particles. The model configurations addressed specific case scenarios and research goals:

(i) *water-saturated, partially saturated, and dry carbonate rocks* (Fig. 2b). These models, with the same configuration applied to the second and fourth layers (North et al., 2023), were used to investigate the effects of water saturation on the thermodynamic properties of a mixture (specific volume, peak shock pressure, shock temperature) under pressures ranging from 4 to 96 GPa of nominal pressure recorded in a nonporous carbonate rock layer (in the range for shock metamorphism, Stöffler et al., 2018). The assumed saturation levels and porosities are summarized in Table 1. In these models, the pores were rounded and non-interconnected, similar to what is often observed in some natural carbonate rocks, particularly in highly crystallized carbonate rocks. We ran models with pure water and pure carbonate rock alongside the multiphase models to compare them to the well understood, pure material thermodynamic properties, replicating the Hugoniot data and validating the models.

(ii) *single pore in a carbonate rock* (Fig. 2c). These models were designed to investigate the effects of saturated or dry pores on the overall response of the rock to weak shock (volumetric strain). The applied pressure was set at 0.10 GPa, and the single pore was positioned to intersect the central axis of the model. The low pressure of 0.10 GPa helped with visualizing volumetric strain and representing material farther from the impact point where fracturing dominates over the effects of shock heat (Stöffler et al., 2018).

(iii) *oriented pore in a carbonate rock* (Fig. 2d). Given that shock wave propagation is radial upon impact, these models were employed to provide a first-order estimate of how shock waves interact with oriented, dry, or water-saturated elongated pores. Probe areas (tracer areas 17 cells across; p1, p2, p3, p4) were used to characterize the shock, in front, behind, or at both ends of a single, elongated, water-filled, or dry, oriented pore (0.144×1.680 mm) at 0° orientation to the shock front. A pressure intensity of 20 GPa was applied to these models. Orientations to the shock front spanned from 0° to 90°, with 10° intervals.

(iv) *simplified fenestral porosity in carbonate rocks* (Fig. 2e; Fig. S1a). Building on the previous model configuration with a single elongated pore, these models expanded the porosity distribution throughout the entire sample to create a reasonable analogy of fenestral porosity, lineation, or bedding in sedimentary rocks. A pressure range of 8–32 GPa was applied to the models. The average initial porosity ranged between 17 and 18% at 0°, 30°, 60°, and 90° orientations between pore-direction and shock wave propagation direction (0° is parallel).

(v) *open fracture and interstitial porosity in carbonate rocks* (Fig. 1f; Fig. S1b). Contrary to previous models, this model featured porosity as interstitial, implemented as a connecting pore space (13% porosity) between rounded grains ranging in size from 136 and 488 μm. Additionally, a horizontal fracture intersected the model, and the pores were either dry or water-filled. The fracture thickness was 120, 280, or 480 μm. A pressure intensity of 20 GPa was applied to each model.

We recorded the maximum peak shock pressures and shock temperatures for each element within the third layer at the end of the simulation. Lagrangian tracers were used to track particle movement and store variables associated with the mobile cells. Volumetric strain was recorded in models featuring either a single large particle (Fig. 1c) or simplified fenestral porosity (lineations, Fig. 1e). Vaporization and escape of water from the rock bulk was not considered, which may lead to underestimation of water's influence on the shock (e.g., fracturing of the rock). Additionally, we did not assess the melting of the carbonate rock as there are no established values for its melting temperature (sources vary up to ~3000 K). Carbonate rocks typically decompose under thermal treatment at ~1000 K (Gunasekaran and Anbalagan, 2007; Olszak-Humienik and Jablonski, 2015) and the calcite mineral may only melt at very high pressures (>40 GPa) and temperatures at the confined pressure (~3000 K, 40 GPa) (Xu et al., 2024). Since applying a cylindrical axis of symmetry did not significantly alter the results (see *Supplementary Material*), we used the axis of symmetry for the models that examined the thermodynamic properties. This allowed us to streamline our simulations without compromising the accuracy of the results. For models containing elongated cracks, we configured a planar shock, as fractures generally adopt a planar-like aspect in carbonate rocks, contrary to round-shaped isolated pores.

*2.3. Normalization of model results*

The velocity of the flyer layer can help normalize mesoscale model results but does not directly translate to the velocity of a meteor striking a planetary surface (impact modelling). To normalize the outputs of these models and better understand the effects of water saturation on carbonate rocks, we recorded the pressure within the flyer layer upon impact onto the underlying layers. We called this the *carbonate rock nominal pressure,* which was directly proportional to the flyer plate velocity used in the models and largely unaffected by the composition of the underlying layers (Fig. 1b; see North et al., 2023).

For the collection of Hugoniot data, we also recorded the peak shock pressure within the sample layer after pressure equalization, but this value was not used for normalization. For instance, if a peak shock pressure of 50 GPa was measured in both nonporous and porous materials, the 50 GPa pressure in the porous material represents a secondary pressure increase that cannot be directly compared to the initial impact conditions in the nonporous material (Wünnemann et al., 2006; Güldemeister et al., 2013; Moreau et al., 2019).

**3. Results**

*3.1. Effects of water saturation on porous carbonate rocks*

*3.1.1. Thermodynamic properties*

Based on water saturation models (Fig. 1a,b, Table 1, b-models), increasing porosity and water content shifted the pressure and specific volume values from those characteristic of pure carbonate rock towards those of pure water (Fig. 2). This shift led to an increase in shock temperatures in carbonate rocks with higher porosity or water content (Fig. 3a). However, the presence of water reduced the overall shock heat distribution across the sample (Fig. 4), as water experienced lower shock temperatures.

In dry porous carbonate rocks, the temperature increase was significantly higher than in water-saturated counterparts. Adding water to the pores damped the temperature increase (Fig. 3a, models with 33% porosity and partial water saturation). The effects of pores and water volumes on the shock temperatures were evident in the bulk (carbonate + water) as well (Fig. 3b). Notably, peak shock pressure values eventually equalized between phases during the shock, as shown in Fig. 4a, where the pressure values of individual phases (water and carbonate rock) converge into a single curve (bulk).

*3.1.2. Pressure and temperature distribution*

Regardless of water particle abundance and the saturation level, water particles consistently stayed cooler than the surrounding carbonate matrix (pressure and temperature maps at ~32 GPa, Fig. 5a,b). The temperature increase in the carbonate rock (Fig. 4) resulted from shock impedance between the water particles and carbonate matrix, redirecting shock waves to deflect into the carbonate rock beneath the less dense water particles (Moreau et al., 2018). The redistribution of heat is similar but more pronounced in dry or partially saturated carbonate rocks (Fig. 5c–f).

In partially saturated carbonate rocks under a nominal 1 GPa shock wave pressure (Fig. 6), the pressure and temperature distribution was notably affected by the arrangement of pores and water particles. Pore collapse played a significant role in heat generation within the carbonate rock compared to the colder water particles. The closure of pores and subsequent pressure spike also lead to an increase of shock temperatures in water particles. At 1 GPa, the impedance contrast between carbonate and locations where water-filled pores were located was minimal and not easily observable (as indicated by blue arrows in Fig. 6g).

*3.1.3. Deformation in single pore models*

In the single pore models (Table 1c, Fig. 1c), both water-filled and dry pores triggered significant fluctuations in volumetric strain during the passage of the shock wave (Fig. 7a,b, left panels). Initially, the carbonate rock experienced compression, leading to negative volumetric strain values (green area in Fig. 7) with relaxation from compression around the pores, owing to changes in impedance (brown area, Fig. 7). This relaxation effect was less pronounced around the water-filled pore. Upon shock wave release, volumetric strain values became positive due to the decompressive (rarefaction) wave propagating through the rock, which initiated tensile stress (pink areas in Fig. 7a,b, right panels). This initiation of rock breakage was also less prominent around the water-filled pore. Positive strain during decompression primarily developed ahead of both dry and water-filled pore, with slight lag behind the dry pore.

*3.2. Elongation and orientation of pores*

*3.2.1. Single elongated pore*

The results for the single oriented and elongated pore model (Table 1, Fig. 1d) capture peak shock pressures and shock temperatures as functions of both pore orientation and water saturation (Fig. 8), along with compiled bulk data. Additionally, time-dependent pressure and temperature increases were recorded at probes p1 and p3 for a pore oriented parallel to the shock front (0° orientation; Fig. 9).

The main findings, regardless of pore orientation, show that water-saturated elongated pores reduced peak shock pressures and temperatures in the carbonate rock compared to empty elongated pores, both above and below the pore (probes p1 and p3) and at the edges (probes p2 and p4) (Fig. 8). Additionally, temperature rise correlates with orientation angle and localized pressure rises. Probe p1, above the pore, peaked in pressure and temperature at a 50° orientation, while edge probes (p2 and p4) peaked at orientations between 80° and 90° (with strong pressure and temperature spikes; Moreau et al., 2018). In contrast, probe p3, located below the pore, showed minimal changes in peak shock pressure or shock temperature relative to orientation. Bulk data indicated a modest temperature rise for orientations between 30° and 50°. The results also suggest that elevated pressures do not necessarily equate to higher temperatures, particularly in probes p1 or p3 at steeper angles (Fig. 8b,c). For probes p1 and p3 on either side of the pore parallel to the shock front, distinct pressure and temperature variations are evident (timewise recordings, Fig. 9). Initially, both probes experienced an immediate pressure spike as the shock front arrived (annot A in Fig. 9b,c). In probe p3, this was followed by an impedance contrast due to the closing pore (annot. B in Fig. 9b) or water. In probe p1, this impedance change led to a sharp pressure drop for the empty pore (Fig. 9b, annot. B) or a partial drop for the water-filled pore (annot. C in Fig. 9b), followed by a quick rapid recovery to the nominal shock pressure (annot. C and D in Fig. 9b). These changes produced secondary temperature increases (Fig. 9c, probe p1: annot. A; probe p3: annot. C or D), which were cumulative (Fig. 9c, probe p1, annot. D). Finally, additional pressure and temperature increases were observed in both probes due to waves interacting from the pore edges (annot. E in Fig. 9b,c).

*3.2.2. Fenestral porosity (lineations)*

Replicating elongated pores in discontinuous patterns (lineations) brought additional insights into shock dynamics and heating (Table 1e, Fig. 1e). A pressure/temperature density distribution of tracers indicates that higher lineation orientation angles (60° and 90°) led to increased overall shock temperatures, accompanied by a decrease in peak shock pressures (Fig. 10, average values for 8 GPa nominal pressure; Fig. 11). At these orientations, the tracer density distribution forms a crescent shape, with excursion of bulk peak shock pressures to higher values with constant temperatures and excursion of shock temperatures to higher values at constant pressures. This highlights the presence of two distinct shock regimes with unique thermodynamic property distributions.

Applying a 2-D distribution of volumetric strain to the models revealed that deformation paths shifted based on the orientation of lineations (Fig. 11). In models with lineations at 0° and 30°, carbonate rock deformation generally followed the original lineation direction, displaying a periodic pattern, reflecting the spacing of the lineations. However, at higher orientations (60° and 90°), the variations in volumetric strain became less pronounced, with the periodicity nearly disappearing in models with 90° lineations. Ultimately, the deformations caused the initial bedding orientation to bend slightly downward relative to the initial impact target surface and impact point (Fig. 11, temperature distribution).

*3.3. Interstitial porosity and fracture collapse*

In the interstitial porosity case study (Fig. 1f, Table 1f), the first indication of the collapse of a virtually added fracture was the pronounced shock temperature contrast above and below the fracture (Fig. 12a,b), which was less intense in the water-saturated porous carbonate rock compared to the dry counterpart. The temperature distribution correlated with the network of interconnecting porosity within the carbonate rock particles. In dry carbonate rock, larger particles exhibited temperature zoning, while in water-saturated rock, hotspots were mainly concentrated in areas with smaller particles. After fracture collapse, the temperature contrast adopted a gradient across the upper carbonate rock layer (Fig. 12c), typically remaining within a few hundred K for the water-saturated rock.

The pressure increase after fracture collapse was more pronounced in water-saturated carbonate rock compared to the dry porous counterpart (Fig. 13a, *ΔP*), contrasting with earlier observations (annot. C and D in Fig. 9b). However, the temperature rise remained higher in the dry porous carbonate rock (Fig. 13b, *ΔT*). During the compression of the water-filled fracture, a slight pressure decrease (1-2 GPa) was noted in the bulk material beneath the fracture. No clear correlation was found between the thickness of the fracture and the increases in pressure or temperature.

A time-based analysis of shock front progression reveals a complex interaction of shock wave reverberations (profiles, Fig. 14), which was not observed in the elongated pore models (Fig. 9). In addition to the expected pressure drop and reflection caused by fracture collapse and the normal shock front progression (annot. A–C in Fig. 14), the intricate porosity network and varying carbonate rock particle sizes produced multiple pressure increases and reverberations within the shock plateau (annot. D and E in Fig. 14, directional arrows). By superimposing longitudinal pressure data over time, we could visually track the reverberation traveling back into the shock plateau (Fig. 14, dashed white arrow intersecting the profiles). Isolating the pressure reverberation from the fracture collapse indicates that it traveled at least 1 mm, ten times the initial fracture width (120 μm).

## 4. Discussion

Water saturation in porous carbonate rocks significantly dampens shock thermodynamic effects, comforting our understanding of groundwater's role in impact dynamics. However, models exploring different types of textural porosity indicate that the shape, distribution, and orientation of the pore space relative to the shock front may significantly influence the final distribution of shock temperatures, pressures, deformations, and the overall response of the target rock to the impact. This introductory study highlights the importance of further research into water saturation in sedimentary rocks with diverse porosity textures. Such research is essential for advancing shock-recovery experiments (Langenhorst and Deutsch, 1994; Langenhorst and Hornemann, 2005), completing the shock classification of sedimentary rocks (Stöffler et al., 2018), and developing more detailed impact simulations on sedimentary targets containing groundwater.

### *4.1. Water-saturated* carbonate rocks

The saturation of carbonate rocks with water significantly influences their shock thermodynamic properties (Figs. 2–4). Under high pressure, the thermodynamic properties of water-saturated carbonate rocks approach those of water, a trend also observed in other water-saturated rocks during planar impact experiments, such as water-saturated sandstone (Arlery et al., 2010; Güldemeister et al., 2013). The effects of water on shock conditions have been extensively studied in laboratory settings (e.g. wet tuff; Anderson and Larson, 1977). Notably, dry porous carbonate rocks exhibit thermodynamic properties similar to those of water-saturated carbonate rocks (Fig. 2), as seen in numerical models with porous and water-saturated sandstones (Güldemeister et al., 2013). However, the temperature rise is ultimately more pronounced in dry porous carbonate rocks (Fig. 3). This suggests that the pressure threshold for decomposition, outgassing (Skála et al., 2002), or melting of porous carbonate rocks may increase when saturated with water, affecting their behavior during crater wall collapse and reducing the amount of decomposed, outgassed, or molten carbonate rocks in the ejecta or on the crater floor. Although we did not consider water vaporization in our models, the reduced shock temperatures in water-saturated host rocks may help explain the deficiency of shock melts in some impact craters, and parallelly further suggesting that water vaporization dynamics could entrain shock melts out of the host rock (Bosumtwi crater, Ghana; Artemieva, 2007).

*4.2. Shock deformation and breakage of porous* carbonate rocks

Carbonate rock deformation is influenced by pore crushing and water impedance contrast (Fig. 7). At low shock pressures, both pore collapse and water saturation contribute to rock breakage. However, water-filled pores reduce deformation, especially at pressures <1 GPa (Fig. 7b). Similar results have been seen in low-pressure dynamic shock experiments on porous, water-saturated reef limestone, where water softens the rock while also enhancing its strength (Cheng et al., 2024).

In parallel, the effect of water on shock wave propagation is similar to the use of water in enhancing rock fracturing during underwater blasts (Wang et al., 2020). Indeed, water increases shock pressure rise from pore collapse and a more homogeneous propagation of the shock wave without a proportional temperature increase (Fig. 13). This effect is seen in water-saturated carbonate rock layers, contributing to more homogeneous rock deformation and fragmentation, as documented in impact craters like Lawn Hill, Australia (Darlington et al., 2016). Such findings suggest the presence of groundwater in sedimentary impact sites, including places like Kaali crater, Estonia (Veski et al., 2007), where water likely played a role in shaping the crater during the impact.

In addition to the localized deformation around dry or water-filled pores, the shock-induced deformation will extend throughout the entire target rock. For lineation models, the angle of the lineations relative to the shock wave influences their reorientation, often aligning with the general slope of the crater floor (Fig. 11) before the host rock eventually experiences an upturn during crater excavation stage (Melosh, 1989). Frictional heating is likely more effective at specific lineation orientations, impacting further the overall shock temperature distribution in the target rock (Spray, 2010), as previously seen (section 3.2, *Elongation and orientation of pores*).

*4.3. Impact cratering in carbonate rocks*

In water-saturated sedimentary rocks that form the upper layers of the target rocks of larger craters (over several kilometers in diameter), chaotic deformations and folding can occur due to vaporization, fluid expansion, and brecciation (Hippertt et al., 2014). This process can lead to the preservation of melt-bearing breccias, often containing remnants of the sedimentary layers with their original or shock-altered laminations, lineations, or even fossil structures (e.g., foraminifera, brachiopods; Garroni, 2023; Newman, 2020). These remnants may provide valuable information on the pre-impact deformation history of the target rock (Fig. 11), although carbonate rocks that compose the upper layers of these large craters may have experienced weaker shock because the shock wave decays rapidly on the surface (Skála, 2002).

In smaller craters dominated by sedimentary rocks, it is less common to find shocked fragments mixed with bedrock or melt products (Gottwald et al., 2020; Kenkmann, 2021). This implies that sedimentary rocks can retain shock imprints without being affected by the additional heat from melt materials. Our study suggests that shock deformation in porous carbonate rocks (such as those with fenestral lineations, interstitial porosity, channels, vuggy pores, and fractures) can provide critical insights into shock dynamics, including the direction and intensity of the shock. This is especially true for heterogeneous carbonate rock formations with strong bedding contrasts, alternating between coherent limestone or dolostone facies and weaker facies like mudstone or sandstone (Lokier and Al Junaibi, 2016, and reference therein; Kentland crater, Weber et al., 2005). These contrasting facies create impedance differences that influence shock wave propagation and heat distribution. Moreover, water saturation within bedded formations is often uneven, affecting the rock response to shock. Impact modeling aims to replicate the results of an impact event as accurately as possible (Collins et al., 2012). This involves defining the target rock using parameters like strength, equations of state, porosity, and thermal softening. Since sedimentary target rocks are highly heterogeneous, modeling impacts on these rocks is challenging. Thus, our work provides information for impact models representing sedimentary rocks, particularly carbonate rocks. It offers new recommendations for those studying specific crater scenarios (e.g., Zhu et al., 2016; Szokaluk et al., 2019; Luther et al., 2023). Incorporating bedding requires adjustments in strength properties, and most importantly, equations of state should be set individually to account for water saturation based on porosity texture. The study also emphasizes the importance of accounting for the radial propagation of shock waves, as bedding or porosity lineations may cause heterogeneous distributions of thermodynamic shock properties, especially shock heating (Fig. 11).

*4.4. Application to other aspects of shock metamorphism*

*4.4.1 Charcoals*

The uneven distribution of shock temperatures can influence the dispersal of shock-heated target fragments into the ejecta blanket. The temperature rise caused by pore crushing on a microscale, which is often overlooked in large-scale impact models, might explain why some craters, such as the Kaali crater (Losiak et al., 2016, 2022), contain charcoal fragments within the ejected material. As seen in our models, significant temperature increases occur around crushed pores (Fig. 5), which also serve as potential breaking points (Fig. 7). This can produce hot fragments with post-shock temperatures capable of charring wood fragments (typically >570 K, Li et al., 2022). However, confirming this hypothesis requires consideration of additional factors influencing ejecta temperatures during atmospheric travel and deposition. These include the heated atmosphere from the blast, vaporization in the ejecta plume (Wrobel et al., 2006; Palumbo and Head, 2018; Carlson et al., 2023), ejecta fragment sizes (Luther et al., 2019), and the thickness of the ejecta blanket.

*4.4.2. Shock classification*

A formal shock classification for carbonate rocks has yet to be developed (Stöffler et al., 2018), especially regarding how shock may affect fossil crystallization and deformation based on rock texture (porosity, microfacies, laminations, and lineation orientation). Shatter cones are currently the most significant but only widely studied marker of shock in carbonate rocks (Baratoux and Melosh, 2003; Baratoux and Reimold, 2016).

The textural porosity, bedding orientation, and deformation of carbonate rocks can also highlight shock effects on a smaller scale, such as the deformation and temperature distribution among clasts (bioclasts or non-bioclasts like ooids; Nichols, 2009). These clasts mineralize with varying textures (e.g., homogeneous crystals, foliation, radial patterns, or microgranular structures). In general, clasts are deposited in a less coherent matrix (especially mudstones and wakestones; Lokier and Al Junaibi, 2016), or can fill up with matrix (e.g., shell organisms, ostracods; Casier et al., 2011), creating significant impedance contrasts (Moreau et al., 2018). Some organisms even exhibit natural undulatory extinction (e.g., trilobites, Gong et al., 2023) which can go in pairs with shock-induced undulatory extinction (Stöffler et al., 2018). These matrix-organism contrasts could lead to specific, organism-dependent shock features, a hypothesis worth exploring in naturally shocked carbonate rocks, shock-recovery experiments, and mesoscale modeling.

**5. Conclusions**

The porosity texture of carbonate rocks and water saturation levels significantly influence shock heating during impact events. Saturated carbonate rocks mitigate shock heating while initiating breakage, deformation, and localized hotspots within the rock. The distribution of pores, such as isolated or interstitial porosity, also affects where the hotspots form, with interstitial porosity and larger grains showing zoned shock heating. Fractures play another crucial role in heat distribution, intensifying heat on one side of the fracture and reheating a previously shock-heated rock after the collapse of the fracture. Shock reverberations further increase cumulative shock temperatures.

We also highlight that textural porosity is important for smaller impacts, especially in rocks with lineations, fenestral textures, or elongated pores and fractures. The radial propagation of the shock wave through the target rock will vary based on the disposition of the porous carbonate rocks, with angles of 50–90° to the shock front having a significant effect on shock heating. Finally, we caution that in impact modeling, bedding must account for porosity, water saturation, and orientation to the shock wave, as all these factors influence deformation and shock heat distribution.

*Acknowledgments* — We gratefully acknowledge the developers of iSALE2D, including Gareth Collins, Kai Wünnemann, Dirk Elbeshausen, Boris Ivanov, and Jay Melosh. Some plots in this work were created with the pySALEPlot tool written by Tom Davison. The iSALEDellen version was used in this work (Collins et al. 2016).

*Funding* — This work was supported by the European Regional Development Fund and the Mobilitas Pluss programme (Grant No. MOBJD639) (2020–2022), and the 2016 Barringer Family Fund for Meteorite Impact Research (Arizona, USA); and National Science Centre, Poland, grant 2020/39/D/ST10/02675.

*Supplementary material* — planar tests and corresponding figures; result table for the thermodynamic properties; additional 2-D plots, graphs, and exemplary figures; strength properties of dolostone.

*Research data* — iSALE configuration files, scripts, plots, and other compilation of data are available as Mendeley Data (doi: 10.17632/85ndv677mj.1 ~~(reserved link) - temporary link, for revision only: [EDITED OUT FOR PRE-PRINT]~~)

*Authors contribution* (CRediT) — **J. Moreau**: project initialization, conceptualization, methodology, validation, formal analysis, investigation, data curation, writing — original draft, writing — review & editing, visualization, funding acquisition; **A. Jõeleht**: conceptualization, methodology, resources, writing — review & editing, supervision, project administration, funding acquisition; **A. Losiak**: project initialization, conceptualization, writing — review & editing; **M.-H. Zhu**: writing — review

& editing; **J. Plado**: conceptualization, methodology, resources, writing — review & editing, supervision, funding acquisition.

**Bibliography**

Amsden A. A., Ruppel H. M., and Hirt C. W. 1980. SALE: A Simplified ALE computer program for fluid flow at all speeds. *Los Alamos National Laboratories Report* LA-8095, Los Alamos, New Mexico. https://doi.org/10.2172/5176006.

Anderson G. D. and Larson D. B. 1977. *Plane shock wave studies of geologic media*. Lawrence Livermore Laboratory, University of California, USA, 23 pp.

Arlery M., Gardou M., Fleureau J. M., and Mariotti C. 2010. Dynamic behaviour of dry and water-saturated sand under planar shock conditions. *Int. J. Impact Eng.* 37(1), 1–10. https://doi.org/10.1016/j.ijimpeng.2009.07.00.

Artemieva N. 2007. Possible reasons of shock melt deficiency in the Bosumtwi drill cores. *Meteorit. Planet. Sci.* 42(4–5), 883–894.

Baranowski P., Kucewicz M., Pytlik M., and Małachowski J. 2022. Shock-induced fracture of dolomite rock in small-scale blast tests. *J. Rock Mech. Geotech. Engineer.* 14(6), 1823–1835. https://doi.org/10.1016/j.jrmge.2021.12.022.

Borg J. P. and Chhabildas L. C. 2011. Three-dimensional dynamic loading simulations of sand. *Air Force Research Laboratory, Munitions Directorate Report*, Eglin AFB, FL, 32542.

Baratoux D. and Melosh H. J. 2003. The formation of shatter cones by shock wave interference during impacting. *Earth Planet. Sci. Lett.* 216(1-2), 43–54. https://doi.org/10.1016/S0012-821X(03)00474-6.

Baratoux D. and Reimold W. U. 2016. The current state of knowledge about shatter cones: Introduction to the special issue. *Meteorit. Planet. Sci.* 51(8), 1389–1434. https://doi.org/10.1111/maps.12678.

Brown W. T. 1989. ANEOS models for shock compression of saturated limestone: Application to ground shock calculation. *Computational Physics & Mechanics Division II, Sandia Report* SAND89-0691, Sandia National Laboratories, New Mexico, 67 pp.

Cabrol N. A. and Grin E. A. 1999. Distribution, classification, and ages of Martian impact crater lakes. *Icarus* 142(1), 160–172.

Carlson M. A., Melosh H. J., and Johnson B. C. 2023. Atmospheric Interactions of Ejecta on Earth and Mars Including the Effect of Vaporization. *Planet. Sci. J.* 4, 194. https://doi.org/10.3847/PSJ/acf9f1.

Casier J. G., Devleeschouwer X., Moreau J., Petitclerc E., and Préat A. 2011. Ostracods, rock facies and magnetic susceptibility records from the stratotype of the Terres d'Haurs Formation (Givetian) at the Mont d'Haurs (Givet, France). *Bull. Inst. R. Sc. N. B.-S.* 81, 97–128.

Cheng B., Huang J., Wang N., Liu T., Li X., and Wang W. 2024. Study on the mechanical properties and fracture characteristics of dry and saturated coral reef limestone under impact loading. *Constr. Build. Mater.* 442, 137618. https://doi.org/10.1016/j.conbuildmat.2024.137618.

Clyde H. M. and William J. W. 2013. Chapter 2 - The Application of the Concepts of Sequence Stratigraphy to Carbonate Rock Sequences, in: *Carbonate Reservoirs, Porosity and Diagenesis in a Sequence Stratigraphic Framework*, edited Clyde H. M., William J. W., *Dev. Sedim.* 67, 23–38, https://doi.org/10.1016/B978-0-444-53831-4.00002-1.

Collins G. S., Melosh H. J., and Ivanov B. A. 2004. Modeling damage and deformation in impact simulations. *Meteorit. Planet. Sci.* 39, 217–231. https://doi.org/10.1111/j.1945-5100.2004.tb00337.x.

Collins G. S., Melosh H. J., and Wünnemann K. 2011. Improvements to the ε-α porous compaction model for simulating impacts into high-porosity solar system objects. *Int. J. Impact Eng.* 38, 434–439. https://doi.org/10.1016/j.ijimpeng.2010.10.013.

Collins G. S., Wünnemann K., Artemieva N., and Pierazzo E. 2012. "Numerical modelling of impact processes." In *Impact Cratering: Processes and Products*, edited by Osinski G. R., and Pierazzo E., Blackwell Publishing Lt, 254–270. https://doi.org/10.1002/9781118447307.ch17.

Collins G. S., Elbeshausen D., Wünnemann K., Davison T. M., Ivanov B., and Melosh H. J. 2016. *iSALE-Dellen manual: A multi-material, multi-rheology shock physics code for simulating impact phenomena in two and three dimensions*. https://doi.org/10.6084/m9.figshare.3473690.

Crawford D. A., Barnouin-Jha O. S., and Cintala M. J. 2003. Mesoscale computational investigation of shocked heterogeneous materials with application to large impact craters. 3rd *International Conference on Large Meteorite Impacts*, abstract #4119, Nördlingen, Germany, CD-ROM.

Darlington V., Blenkinsop T., Dirks P., Salisbury J., and Tomkins A. 2016. The Lawn Hill annulus: An Ordovician meteorite impact into water-saturated dolomite. *Meteorit. Planet. Sci.* 51, 2416–2440. https://doi.org/10.1111/maps.12734.

Davison T. M., Collins G. S., and Bland P. A. 2016. Mesoscale Modeling of Impact Compaction of Primitive Solar System Solids. *Astrophys. J.* 821, 17 pp. https://doi.org/10.3847/0004-637X/821/1/68.

Davison T. M., Derrick J. G., Collins G. S., Bland P. A., Rutherford M. E., Chapman D. J., and Eakins D. E. 2017. Impact-induced compaction of primitive solar system solids: The need for mesoscale modelling and experiments. *Procedia Engineer.* 204, 405–412. https://doi.org/10.1016/j.proeng.2017.09.801.

Demurtas M., Smith S. A. E., Spagnuolo E., and Di Toro G. 2021. Frictional properties and microstructural evolution of dry and wet calcite–dolomite gouges. *Solid Earth* 12, 595–612. https://doi.org/10.5194/se-12-595-2021.

Dufresne A., Poelchau M. H., Kenkmann T., Deutsch A., Hoerth T., Schäfer F., and Thoma K. 2013. Crater morphology in sandstone targets: The MEMIN impact parameter study. *Meteorit. Planet. Sci.* 48(1), 50–70. https://doi.org/10.1111/maps.12024.

Dunham R. J. 1962. Classification of carbonate rocks according to depositional texture. In: *Classification of Carbonate Rocks*, ed. W.E. Ham, *Am. Assoc. Petr. Geol. Memoir* 1, 108–121.

Eppelbaum L., Kutasov I., and Pilchin A. 2014. Applied Geothermics. Springer, Heidelberg, New York, Dordrecht, London.

Ford D. C. and Williams P. 2007. *Karst Hydrogeology and Geomorphology*, John Wiley and Sons Ltd., 562 pp.

Garroni N. D. 2023. *The Fate of Carbonate Rocks During Hypervelocity Impacts: Case Studies from Three Impact Structures on Earth*. Electronic Thesis and Dissertation Repository, 9129. https://ir.lib.uwo.ca/etd/9129.

Gong L., Gao X., Stow D., Qu F., Zhang G., and Liu P. 2023. What continued after the mass extinction: insights from carbonate microfacies and biological evolution around the Permian–Triassic boundary in the middle Upper Yangtze Platform, SW China. *Geol. Mag.* 160, 35–59. . https://doi.org/10.1017/S0016756822000632.

Gottwald M., Kenkmann T., and Reimold W. U. 2020. *Terrestrial impact structures. The TanDEM-X atlas*. Munich: Verlag Dr. Friedrich Pfeil. 608 p.

Gunasekaran S. and Anbalagan G. 2007. Thermal decomposition of natural dolomite. *Bull. Mater. Sci.* 30(4), 339–344.

Güldemeister N., Wünnemann K., Durr N., and Hiermaier S. 2013. Propagation of impact-induced shock waves in porous sandstone using mesoscale modeling. *Meteorit. Planet. Sci.* 48, 115–133. https://doi.org/10.1111/j.1945-5100.2012.01430.x.

Herrick R. R., Bateman E. M., Crumpacker W. G., and Bates D. 2018. Observations from a global database of impact craters on Mercury with diameters greater than 5 km. *J. Geophys. Res.-Planet.* 123(8), 2089–2109.

Hippertt J. P., Lana C., Weinberg R. F., Tohver E., Schmieder M., Scholz R., Gonçalves L., and Hippertt J. F. 2014. Liquefaction of sedimentary rocks during impact crater development. *Earth Planet. Sc. Lett.* 408, 285–295. https://doi.org/10.1016/j.epsl.2014.09.045.

Hoerth T., Schäfer F., Thoma K., Kenkmann T., Poelchau M. H., Lexow B., and Deutsch A. 2013. Hypervelocity impacts on dry and wet sandstone: Observations of ejecta dynamics and crater growth. *Meteorit. Planet. Sci.* 48(1), 23–32. https://doi.org/10.1111/maps.12044.

Hörz F., Cintala M. J., Thomas-Keprta K. L., Ross D. K., and Clemett S. J. 2019. Unconfined shock experiments: A pilot study into the shock-induced melting and devolatilization of calcite. *Meteorit. Planet. Sci.* 55, 102–129. https://doi.org/10.1111/maps.13424.

Ivanov B. A. 2005. Shock Melting of Permafrost on Mars: Water Ice Multiphase Equation of State for Numerical Modeling and Its Testing. *36th Lunar and Planetary Science Conference*, abstract #1232, League City, Texas.

Ivanov B. A., Deniem D., and Neukum G. 1997. Implementation of dynamic strength models into 2D hydrocodes: Applications for atmospheric breakup and impact cratering. *Int. J. Impact Eng.* 20, 411–430. https://doi.org/10.1016/S0734-743X(97)87511-2.

Janjuhah H. T., Kontakiotis G., Wahid A., Khan D. M., Zarkogiannis S. D., and Antonarakou A. 2021. Integrated porosity classification and quantification scheme for enhanced carbonate reservoir quality: Implications from the Miocene Malaysian carbonates. *J. Mar. Sci. Eng.* 9, 1410. https://doi.org/10.3390/jmse9121410.

Kenkmann T. 2021. The terrestrial impact crater record: A statistical analysis of morphologies, structures, ages,

lithologies, and more. *Meteorit. Planet. Sci.* 56(5), 1024–1070. https://doi.org/10.1111/maps.13657.

Kenkmann T., Sundell K. A., and Cook D. 2018. Evidence for a large Paleozoic impact crater strewn field in the rocky mountains. *Sci. Rep.-UK* 8, 13246, 1–12. https://doi.org/10.1038/*s41598-018-31655-4*.

Kurosawa K., Ono H., Niihara T., Sakaiya T., Kondo T., Tomioka N., Mikouchi T., Genda H., Matsuzaki T., Kayama M., Koike M., Sano Y., Murayama M., Satake W., and Matsui T. 2022. Shock recovery with decaying compressive pulses: Shock effects in calcite ($CaCO_3$. around the Hugoniot elastic limit. *J. Geophys. Res.-Planets* 127, e2021JE007133, 1–15.

Langenhorst F. and Deutsch A. 1994. Shock Experiments on Pre-Heated Alpha-Quartz and Beta-Quartz: I. Optical and Density Data. *Earth Planet. Sc. Lett.* 125, 407–420. https://doi.org/10.1016/0012-821x(94)90229-1.

Langenhorst F. and Hornemann U. 2005. Shock experiments on minerals: Basic physics and techniques. *EMU Notes in Mineralogy* 7, 357–387.

Larsen G. and Chilingor G. V. 1979. Chapter 1 Introduction-Diagenesis of Sediments and Rocks, in: *Diagenesis in Sediments and Sedimentary Rocks*, 579 pp., *Dev. Sedim.* 25(A), 1–29.

Li G., Gao L., Liu F, Qiu M., and Dong G. 2022. Quantitative studies on charcoalification: Physical and chemical changes of charring wood. *Fund. Res.* 4(1), 113–122. https://doi.org/10.1016/j.fmre.2022.05.014.

Littlefield D. L., Bauman P. T., and Molineux A. 2007. Analysis of formation of the Odessa crater, *Int. J. Impact Eng.* 34(12), 1953–1961, https://doi.org/10.1016/j.ijimpeng.2006.12.005.

Lokier S. W. and Al Junaibi M. 2016. The petrographic description of carbonate facies: are we all speaking the same language? *Sedimentology* 63, 1843–1885. https://doi.org/10.1111/sed.12293.

Losiak A., Wild E. M., Geppert W. D., Huber M. S., Jõeleht A., Kriiska A., Kulkov A., Paavel K., Pirkovic I., Plado J., Steier P., Välja R., Wilk J., Wisniowski T., and Zanetti M. 2016. Dating a small impact crater: an age of Kaali crater (Estonia. based on charcoal emplaced within proximal ejecta. *Meteorit. Planet. Sci.* 51, 681–695. https://doi.org/10.1111/maps.12616.

Losiak A., Belcher C. M., Plado J., Jõeleht A., Herd C. D. K., Kofman R. S., Szokaluk M., Szczuciński W., Muszyński A., Wild E. M., and Baker S. J. 2022. Small impact cratering processes produce distinctive charcoal assemblages. *Geology* 50(11), 1276–1280. https://doi.org/10.1130/G50056.1.

Lucia F. J. 2004. "Origin and petrophysics of dolostone pore space." In *The Geometry and Petrogenesis of Dolomite Hydrocarbon Reservoirs*, edited by Braithwaite C. J. R., Rizzi, G. & Darke, Geological Society, London, Special Publications, 235, 141–155.

Luther R., Artemieva N., and Wünnemann K. 2019. The effect of atmospheric interaction on impact ejecta dynamics and deposition. *Icarus* 333, 71–86. https://doi.org/10.1016/j.icarus.2019.05.007.

Luther R., Schmalen A., and Artemieva N. 2023. Campo del Cielo modeling and comparison with observations: II. Funnels and craters. *Meteorit. Planet. Sci.* 58(12), 1832–1847. https://doi: 10.1111/maps.141041Ó2023.

Marchi S., Ermakov A. I., Raymond C. A., Fu R. R., O'Brien D. P., Bland M. T., Ammannito E., De Sanctis M. C., Bowling T., Schenk P., Scully J. E. C., Buczkowski D. L., Williams D. A., Hiesinger H., and Russell C. T. 2016. The missing large impact craters on Ceres. *Nat. Commun.* 7, 12257. https://doi.org/10.1038/ncomms12257.

Martinez I., Deutsch A., Schärer U., Ildefonse P., Guyot F., and Agrinier P. 1995. Shock recovery experiments on dolomite and thermodynamical calculations of impact induced decarbonation, *J. Geophys. Res.* 100(B8), 15465–15476. https://doi.org/10.1029/95JB01151.

Melosh H. J. 1989. Impact Cratering: A Geologic Process. New York: Oxford Univ Press, 253 pp.

Melosh H. J. J., Ryan E. V. V., and Asphaug E. 1992. Dynamic fragmentation in impacts: Hydrocode simulation of laboratory impacts. *J. Geophys. Res.* 97, 14735–14759. https://doi.org/10.1029/92JE01632.

Montmerle T., Augereau J.–C., Chaussidon M., Counelle M., Marty B., and Morbidelli A. 2006. 3. Solar System formation and early evolution: the first 100 million year". *Earth Moon Planets* 98(1–4), 39–95.

Moreau J., and Schwinger S. 2020. Heat diffusion in numerically shocked ordinary chondrites and its contribution to shock melting. *Phys. Earth Plan. In.* 310, 106630. https://doi.org/10.1016/j.pepi.2020.106630.

Moreau J., Kohout T., and Wünnemann K. 2017. Shock-darkening in ordinary chondrites: Determination of the pressure-temperature conditions by shock physics mesoscale modeling. *Meteorit. Planet. Sci.* 52, 2375–2390. https://doi.org/10.1111/maps.12935.

Moreau J., Kohout T., and Wünnemann K. 2018. Melting efficiency of troilite-iron assemblages in shock-darkening: Insight from numerical modeling. *Phys. Earth Plan. In.* 282, 25–38. https://doi.org/10.1016/j.pepi.2018.06.006.

Moreau J., Kohout T., Wünnemann K., Halodova P., and Haloda J. 2019. Shock physics mesoscale modeling of shock stage 5 and 6 in ordinary and enstatite chondrites. *Icarus* 332, 50–65.

Newman J. D. 2020. *Impact-Generated Dykes and Shocked Carbonates from the Tunnunik and Haughton Impact Structures, Canadian High Arctic*. Electronic Thesis and Dissertation Repository, 6950. https://ir.lib.uwo.ca/etd/6950.

Nichols G. 2009. *Sedimentology and Stratigraphy*. Blackwell Science Ltd., London, 335 pp.

North T. L., Collins G. S., Davison T. M., Muxworthy A. R., Steele S. C., and Fu R. R. 2023. The heterogeneous response of martian meteorite Allan Hills 84001 to planar shock. *Icarus* 390, 115322, 1–16. https://doi.org/10.1016/j.icarus.2022.115322.

Osinski G. R., Grieve R. A. F., Ferrière L., Losiak A., Pickersgill A. E., Cavosie A. J., Hibbard S. M., Hill P. J. A., Bermudez J. J., Marion C. L., Newman J. D., and Simpson S. L. 2022. Impact Earth: A review of the terrestrial impact record. *Earth-Sci. Rev.* 232, 104112. https://doi.org/10.1016/j.earscirev.2022.104112.

Olszak-Humienik M. and Jablonski M. 2015. Thermal behavior of natural dolomite. *J. Therm. Anal. Calorim.* 119, 2239–2248. https://doi.org/10.1007/s10973-014-4301-6.

Ott R., Gallen S. F., and Helman D. 2023. Erosion and weathering in carbonate regions reveal climatic and tectonic drivers of carbonate landscape evolution. *Earth Surf. Dynam.* 11, 247–257. https://doi.org/10.5194/esurf-11-247-2023, 2023.

Palumbo A. M. and Head J. W. 2018. Impact cratering as a cause of climate change, surface alteration, and resurfacing during the early history of Mars. *Meteorit. Planet. Sci.* 53(4), 687–725. https://doi.org/10.1111/maps.13001.

Patel M. B. and Shah M. V. 2015. Strength characteristics for limestone and dolomite rock matrix using tri-Axial system. *Int. J. Eng., Sci. Tech.* 1(11), 2349–784X, 114-124.

Plado J. 2012. Meteorite impact craters and possibly impact-related structures in Estonia. *Meteorit. Planet. Sci.* 47(10), 1590–1605. https://doi.org/10.1111/j.1945-5100.2012.01422.x.

Plummer P. S. 2021. The Neoproterozoic Gillen Formation, Amadeus Basin, central Australia: an intra-salt petroleum system and viable exploration target? *The APPEA Journal* 61, 236–252. https://doi.org/10.1071/AJ20040.

Riedel W., Wicklein M., and Thoma K. 2008. Shock properties of conventional and high strength concrete: Experimental and mesomechanical analysis. *Int. J. Impact Eng.* 35, 155–171. https://doi.org/10.1016/j.ijimpeng.2007.02.001.

Robbins S. J. 2018. A New global Database of lunar impact craters >1–2 km: 1. Crater locations and sizes, comparisons with published databases, and global analysis. *J. Geophys. Res.-Planets* 124(4), 871–892.

Rosentau A., Klemann V., Bennike O., Steffen H., Wehr J., Latinović M., Bagge M., Ojala A., Berglund M., Becher G. P., Schoning K., Hansson A., Nielsen L., Clemmensen L. B., Hede M. U., Kroon A., Pejrup M., Sander L., Stattegger K., Schwarzer K., Lampe R., Lampe M., Uścinowicz S., Bitinas A., Grudzinska I., Vassiljev J., Nirgi T., Kublitskiy Y., and Subetto D. 2021. A Holocene relative sea-level database for the Baltic Sea. *Quaternary Sci. Rev.* 266, 107071.

Scuderi M. M., Niemeijer A. R., Collettini C., and Marone C. 2013. Frictional properties and slip stability of active faults within carbonate–evaporite sequences: The role of dolomite and anhydrite. *Earth Planet. Sc. Lett.* 369–370, 220–232.

Skála R. 2002. Shock-induced phenomena in limestones in the quarry near Ronheim, the Ries Crater, Germany. *Bull. Czech Geol. Surv.* 77(4), 313–320.

Skála R., Ederová J., Matějka P., and Hörz F. 2002. Mineralogical investigations of experimentally shocked dolomite: Implications for the outgassing of carbonates. In *Catastrophic events and mass extinctions: Impacts and Beyond*, edited by Koeberl C. and MacLeod K. G., Geological Society of America, USA.

Spray J. G. 2010. Frictional melting processes in planetary materials: from hyperve-locity impact to earthquakes. *Annu. Rev. Earth Pl. Sc.* 38, 221–254.

Stöffler D., Hamann C., and Metzler K. 2018. Shock metamorphism of planetary silicate rocks and sediments: Proposal for an updated classification system. *Meteorit. Planet. Sci.* 53, 5–49. https://doi.org/10.1111/maps.12912.

Szokaluk M., Jagodziński R., Muszyński A., and Szczuciński W. 2019. Geology of the Morasko craters, Poznań, Poland—Small impact craters in unconsolidated sediments. *Meteorit. Planet. Sci.* 54, 1478–1494. https://doi.org/10.1111/maps.13290.

Takita H. and Sumita I. 2013. Low-velocity impact cratering experiments in a wet sand target. *Phys. Rev.* 88, 022203, 1–10.

Veski S., Heinsalu A., Poska A., Saarse L., and Vassiljev J. 2007. "The physical and social effects of the Kaali meteorite impact—A review." In *Comet/asteroid impacts and human society*, edited by Bobrowsky P. T., and Rickman H., Berlin: Springer-Verlag., 265–275.

Wang J., Yang J. Wu, F., Hu T., and Al Faisal S. 2020. Analysis of fracture mechanism for surrounding rock hole based on water-filled blasting. *International J. Coal Sci. Tech.* 7, 704–713. https://doi.org/10.1007/s40789-020-00327-y.

Weber J. C., Poulos C., Donelick R. A., Pope M. C., and Heller, N. 2005. The Kentland Impact Crater, Indiana (USA): An Apatite Fission-Track Age Determination Attempt. In: *Impact Tectonics. Impact Studies*, edited Koeberl C., Henkel H., Springer, Berlin, Heidelberg, 447–466. https://doi.org/10.1007/3-540-27548-7_18.

Wrobel K., Schultz P., and Crawford D. 2006. An atmospheric blast/thermal model for the formation of high-latitude pedestal craters. *Meteorit. Planet. Sci.* 41(10), 1539–1550.

Wünnemann K., Collins G. S., and Melosh H. J. 2006. A strain-based porosity model for use in hydrocode simulations of impacts and implications for transient crater growth in porous targets. *Icarus* 180, 514–527. https://doi.org/10.1016/j.icarus.2005.10.013.

Xu Z.-X., Xu F.-Y., Liu Y., Geng H., Li Z.-G., and Hu J. 2024. High-pressure melting behaviors of calcite from first-principles simulation. *Physica B-Condensed Matter* 680, 415810, pp. 7.

Zhu M.-H., Bronikowska M., and Losiak A. 2016. The formation of Kaali crater, Estonia: Insights from numerical modeling. *79th Annual Meeting of the Meteoritical Society*, abstract #6325, Berlin, Germany.

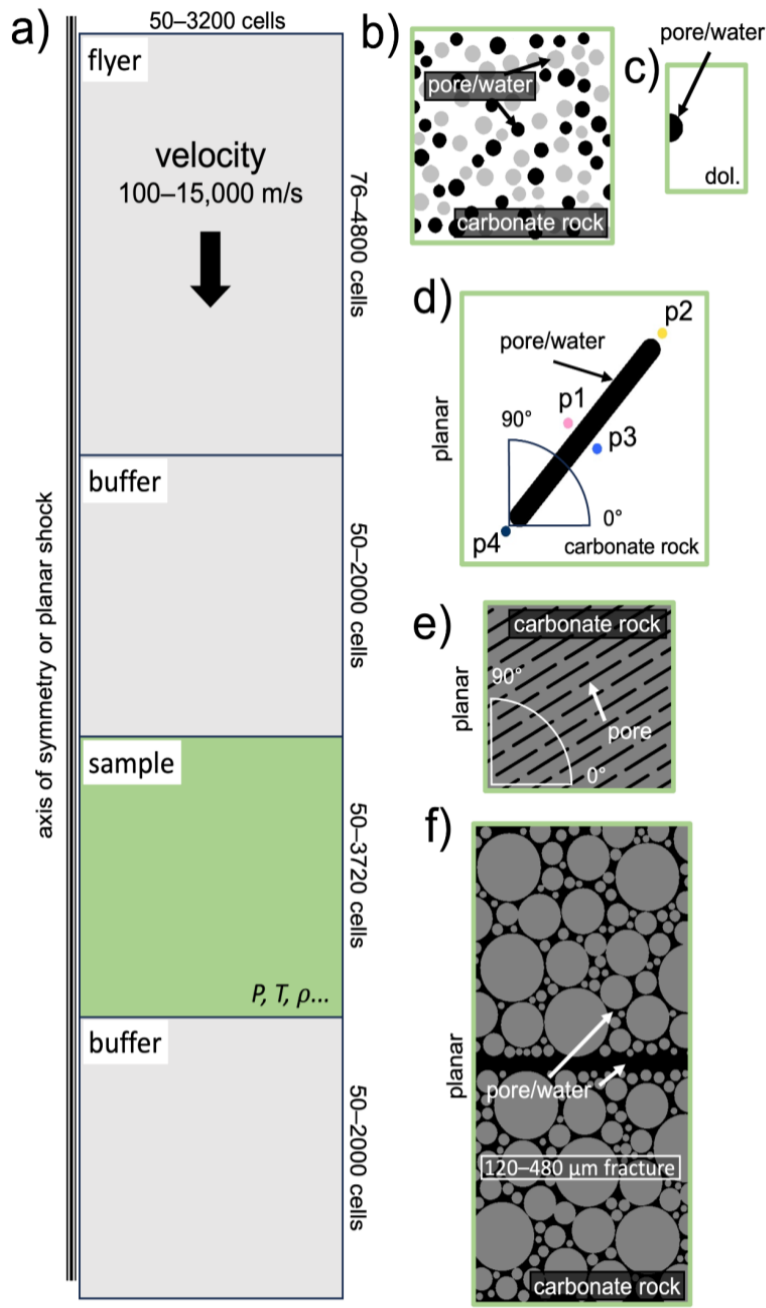

**Fig. 1.** Mesoscale model setup for water saturation in texturally porous carbonate rocks with a) general configuration of all mesoscale models following their composition, resolution, and parametrization requirements shown in b–f) and Table 1 (incl. sizes). Model variables are assessed within the sample layer, and they are $P$ — peak shock pressure, $T$ — shock temperature, $\rho$ — density. Volumetric strain is assessed in the sample of the model shown in c) and e). The "nominal pressure" is assessed within the flyer plate of a pure carbonate rock model similar to a). Boundary conditions are as follows: top/bottom: *outflow*, left/right: *free-slip*. A bitmap file is often used to define the sample and buffers in b–f). Cylindrical symmetry is applied to models shown in b) and c). In d), probes p1–p4 are clusters of tracers used to record variables at different positions along the pore.

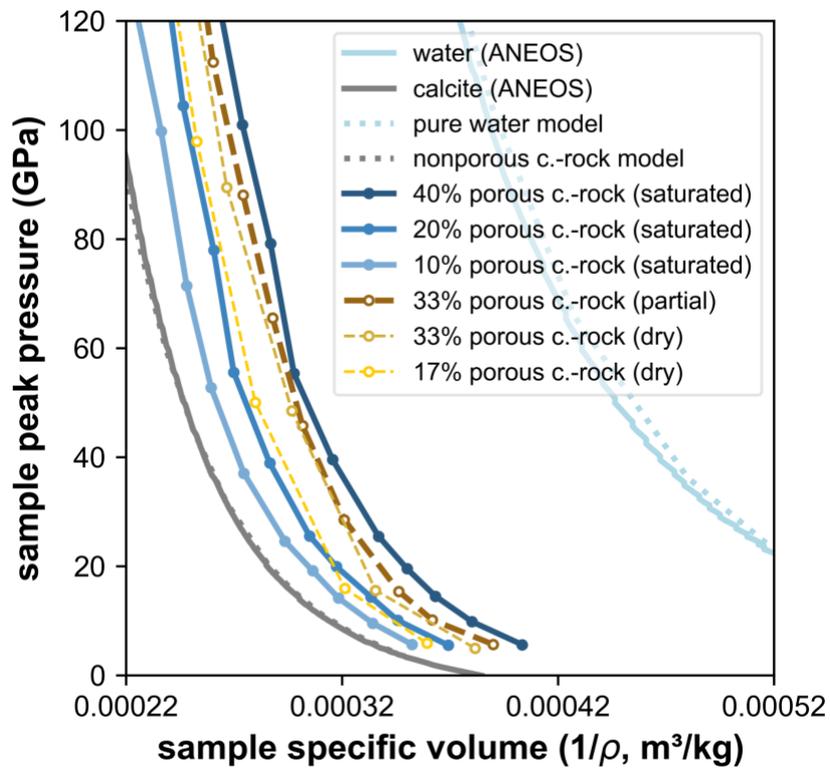

**Fig. 2.** Hugoniot data of numerical models (similar to Fig. 1b) for water saturation in carbonate rocks, giving values for the entire sample layer peak pressure as a function of specific volume. ANEOS data is directly translated from iSALE setup files whereas 100% carbonate rock or 100% water values are estimated from models, instead. All models are composed of carbonate rock, apart from the 100% water model. ANEOS is used for all materials. Calcite ANEOS is used to simulate the carbonate rock. Partial saturation: 18% water-filled pores and 15% empty pores. *c.-rock*: carbonate rock.

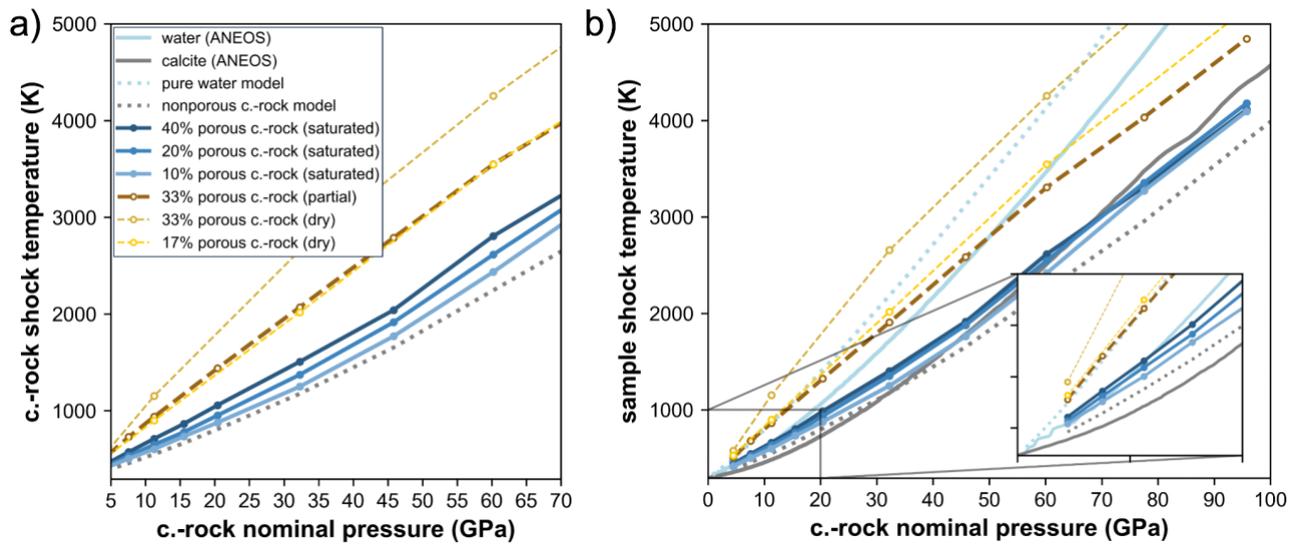

**Fig. 3.** Influence of water saturation on the carbonate rock phase and sample bulk (carbonate + water, if present) from numerical models (similar to Fig. 1b) with a) carbonate rock shock temperature as a function of carbonate rock nominal pressure and b) sample bulk shock temperature as a function of carbonate rock nominal pressure. ANEOS is used for all materials. Calcite ANEOS is used to simulate the carbonate rock. Nominal pressure: pressure recorded in the carbonate rock flyer plate in a model devoid of water/pore particles. Partial saturation: 18% water-filled pores and 15% empty pores. *c.-rock*: carbonate rock.

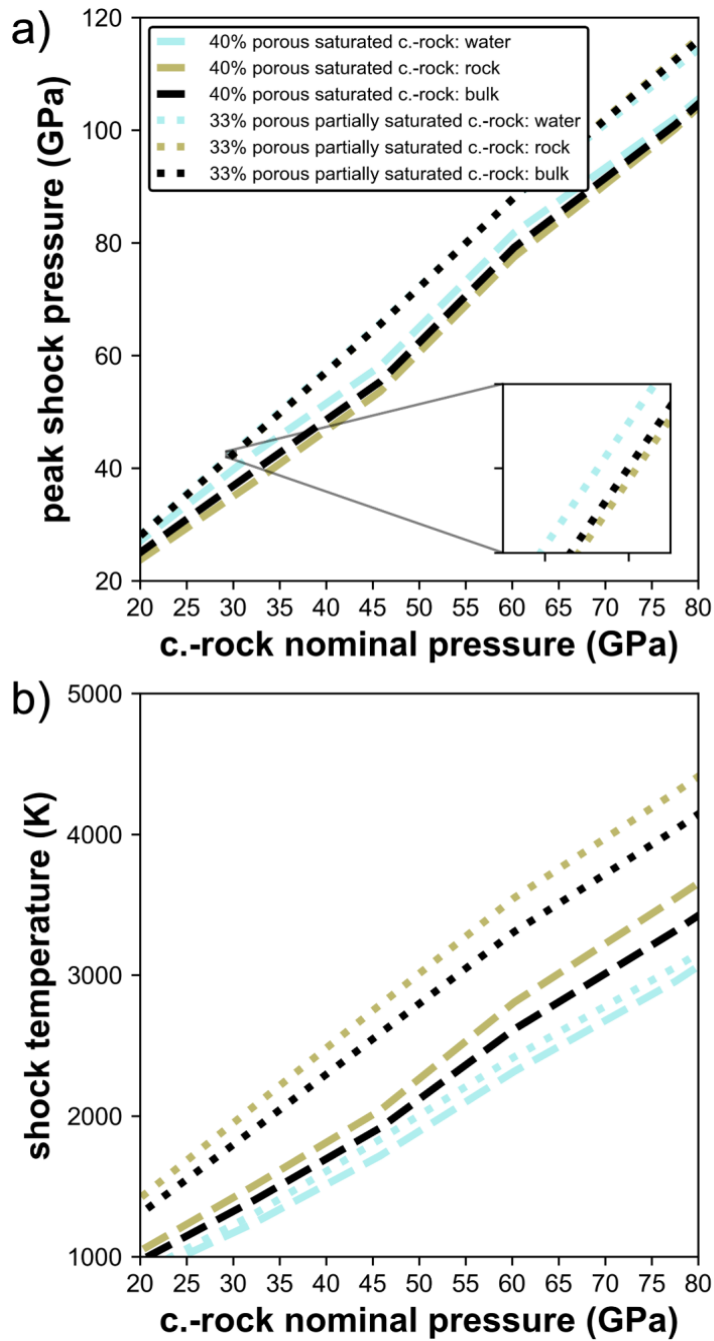

**Fig. 4.** Distribution of shock pressures and shock temperatures between the carbonate rock and water upon shock in two chosen numerical models (similar to Fig. 1b), with a) peak shock pressure as a function of carbonate rock nominal pressure and b) shock temperature as a function of carbonate rock nominal pressure. Values are also shown for the mixtures (carbonate rock + water). Inset in a) demonstrates that pressures equalize through the bulk mixture. Nominal pressure: pressure recorded in the carbonate rock flyer plate in a model devoid of water/pore particles. Partial saturation: 18% water-filled pores and 15% empty pores. *c.-rock*: carbonate rock.

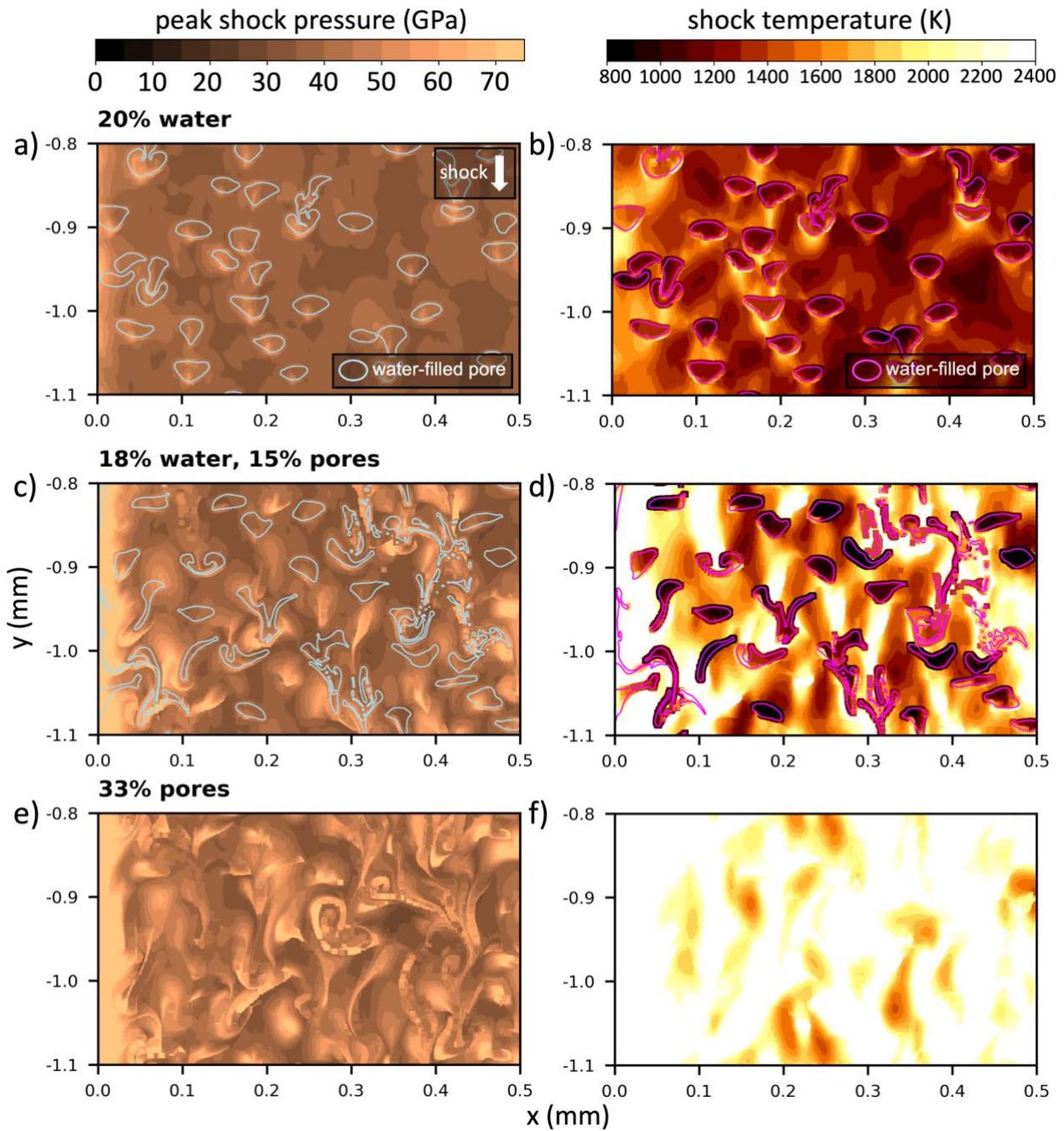

**Fig. 5.** Effect of water saturation on the distribution of peak shock pressures and shock temperatures in the carbonate rock at high shock pressure (~32 GPa nominal pressure) in a–b) water-saturated, c–d) partially water-saturated, and e–f) dry porous carbonate rock (based on models similar to Fig. 1b). Location of water-filled pores (20 cells per particle radius) is indicated. The initial shock front direction is indicated, and models are shown right after passage of the shock wave.

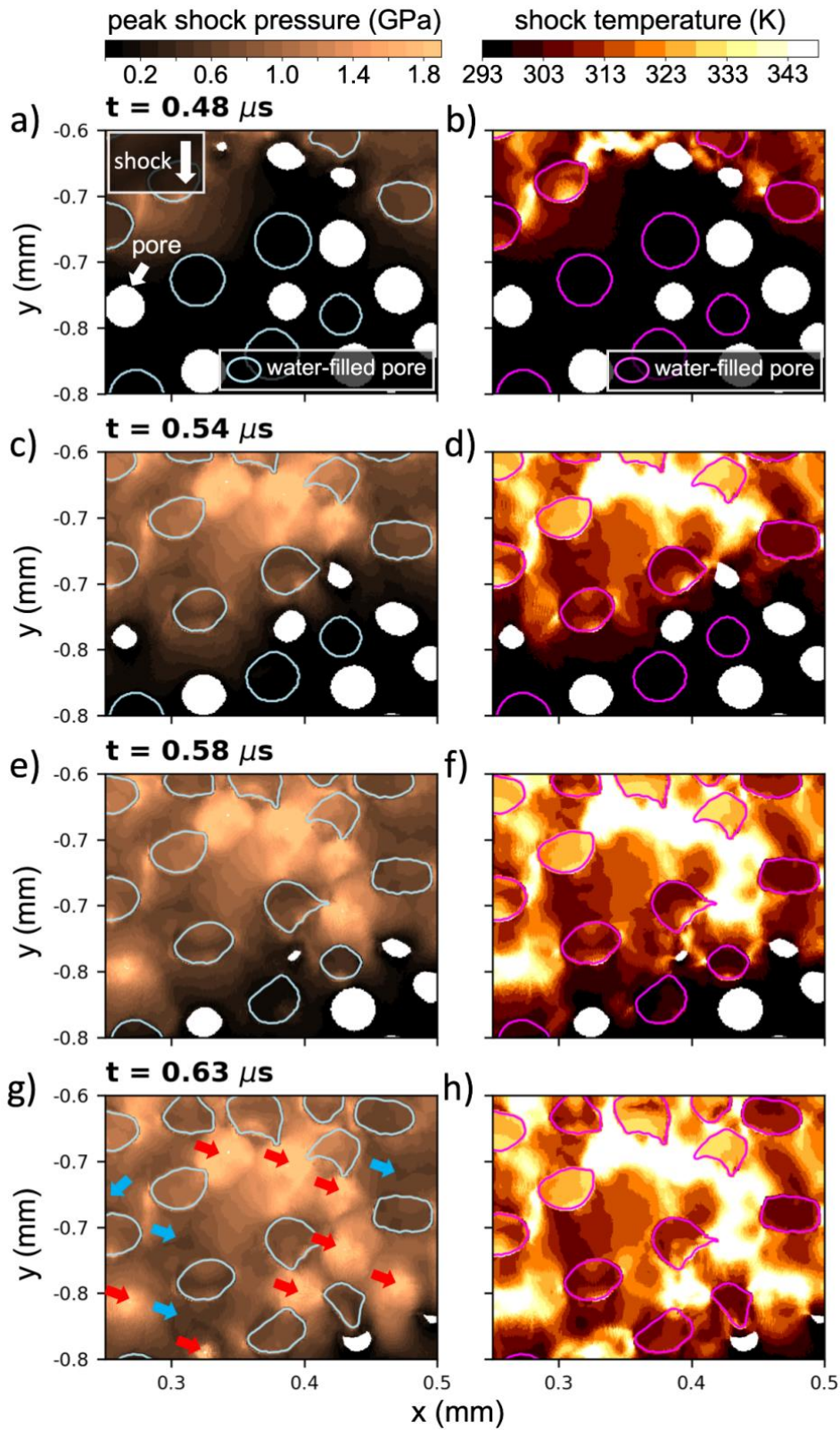

**Fig. 6.** Progression of the shock wave in a partially saturated carbonate rock at low pressure (at ~1 GPa nominal pressure), in a selected area and at selected time intervals of the model (Fig. 1b). Red arrows indicate hotspots from pore crushing and blue arrows indicate "coldspots". Location of water particles (20 cells per particle radius) is indicated. Pores (20 cells per particle radius) remain visible (white) before they close completely from shock loading. The initial shock front direction is indicated, and models are shown right after passage of the shock wave.

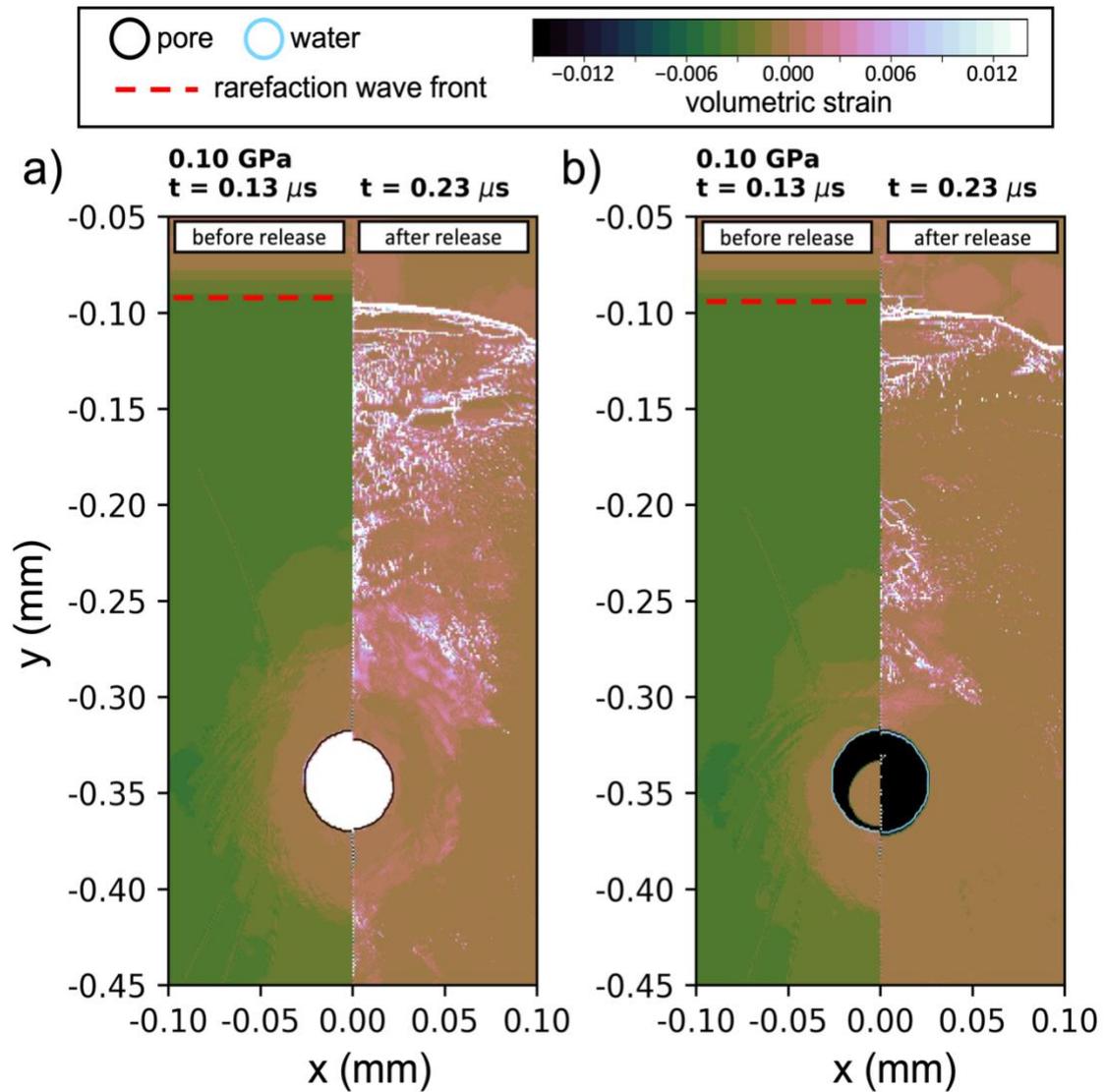

**Fig. 7.** 2-D distribution maps of volumetrics strain during and after the passage of the compressive wave (0.10 GPa) through a) a dry pore and b) a water-filled pore in the carbonate rock (Fig. 1c). The negative or positive values for volumetric strain are equivalent to the compressive or tensile stresses at release of the shock, respectively. Pore resolution is 27 cells per particle radius.

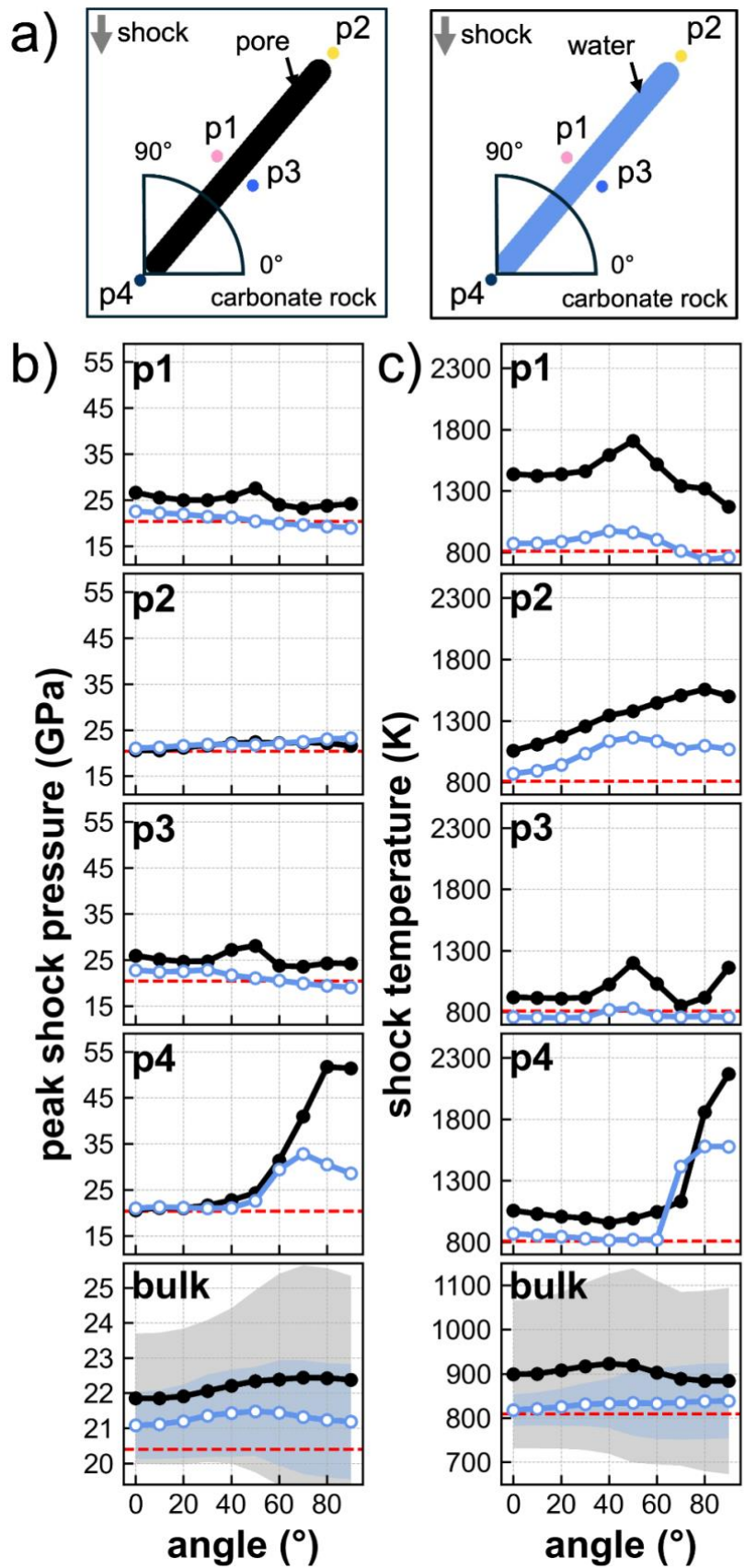

**Fig. 8.** Effect of the orientation of an elongated pore to the shock wave front on peak shock pressures and peak shock temperatures in a carbonate rock (Fig. 1d). a) Orientations ranged from 0° to 90°, and the model either included an empty or water-saturated elongated pore. b,c) Results are extracted from four probe locations (p1, p2, p3, p4) situated at pore extremities and pore walls: b) recorded peak shock pressures and c) recorded peak shock temperatures in each probe and the bulk (shaded areas are standard deviations). Red dashed lines are the nominal, or reference shock pressures and shock temperatures in the carbonate rock. The direction of the shock front is indicated in (a).

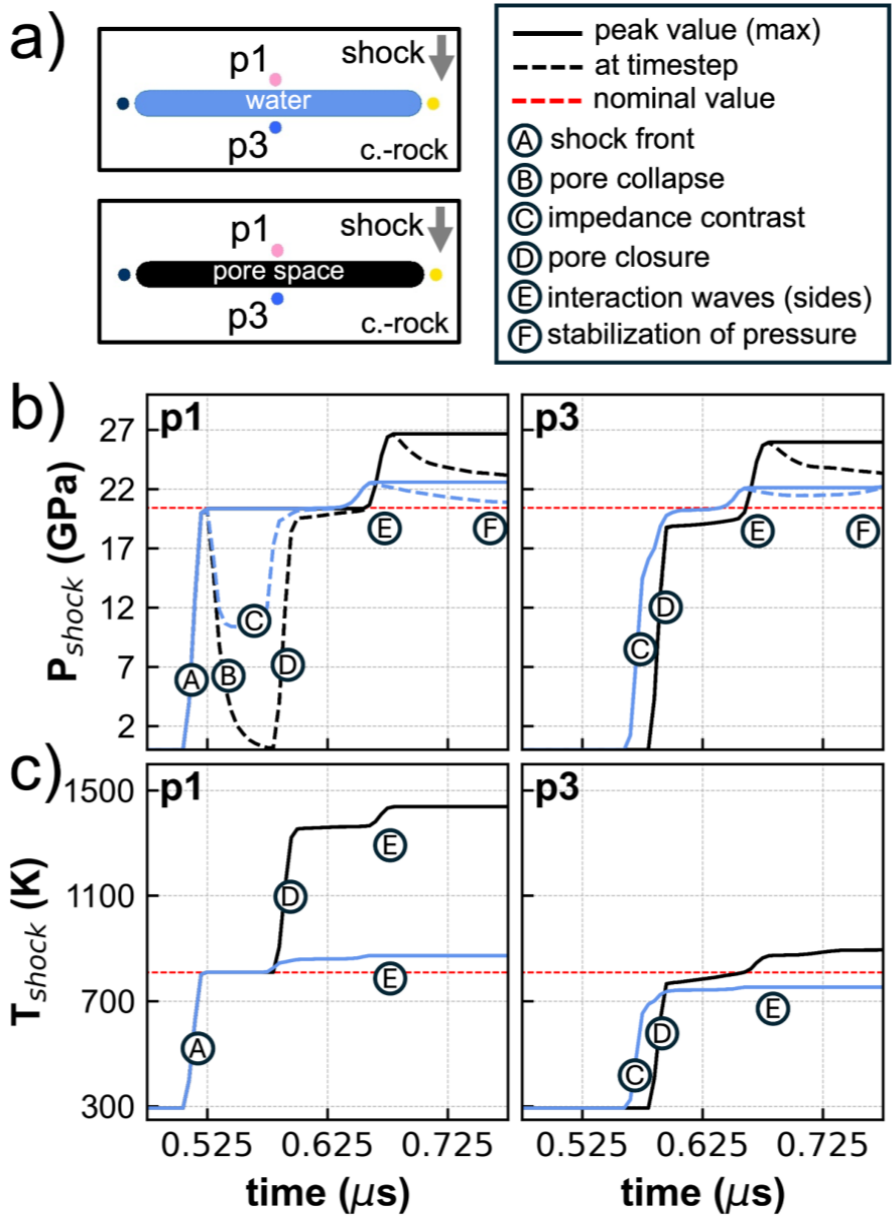

**Fig. 9.** Time-wise pressure and temperature recordings in probes p1 and p3 situated on either side of an empty or water-saturated pore in the carbonate rock (Fig. 1d): a) model depictions of the pores and probes with added legend; b) recorded shock pressure and c) shock temperature over time in the corresponding probes. Black or blue dashed lines are the pressure and temperature at the time of recording and plain lines are the maximal, or peak values recorded along the simulation. Red dashed lines are the nominal, or reference pressures and temperatures in the carbonate rock. Numbering illustrates different effects of shock wave propagation. Interaction waves (E) are shock waves produced at the pore extremities. The direction of the shock front is indicated in (a). *c.-rock*: carbonate rock.

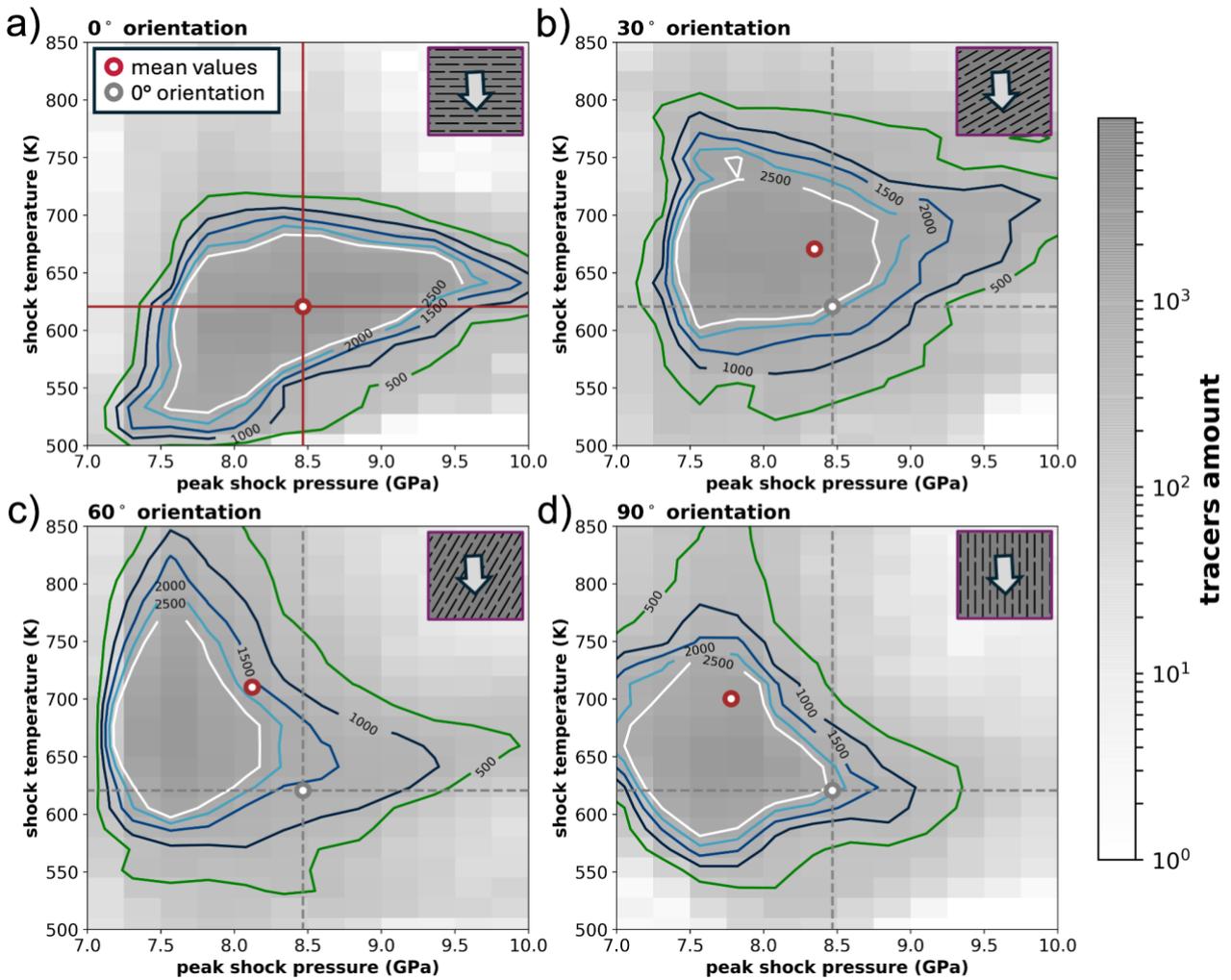

**Fig. 10.** Shock pressure and temperature density distributions of tracers within fenestral porosity models (lineations in carbonate rocks, Fig. 1e) at different degrees of orientation to the shock wave front. In b–d) panels, the crosshair dashed lines indicate the mean values for the model at 0° of orientation (a), therefore highlighting the corresponding variations in pressure and temperature as a function of lineation orientation. Model insets, with shock front direction, and levels corresponding to the amount of tracers in the density maps are added for clarity.

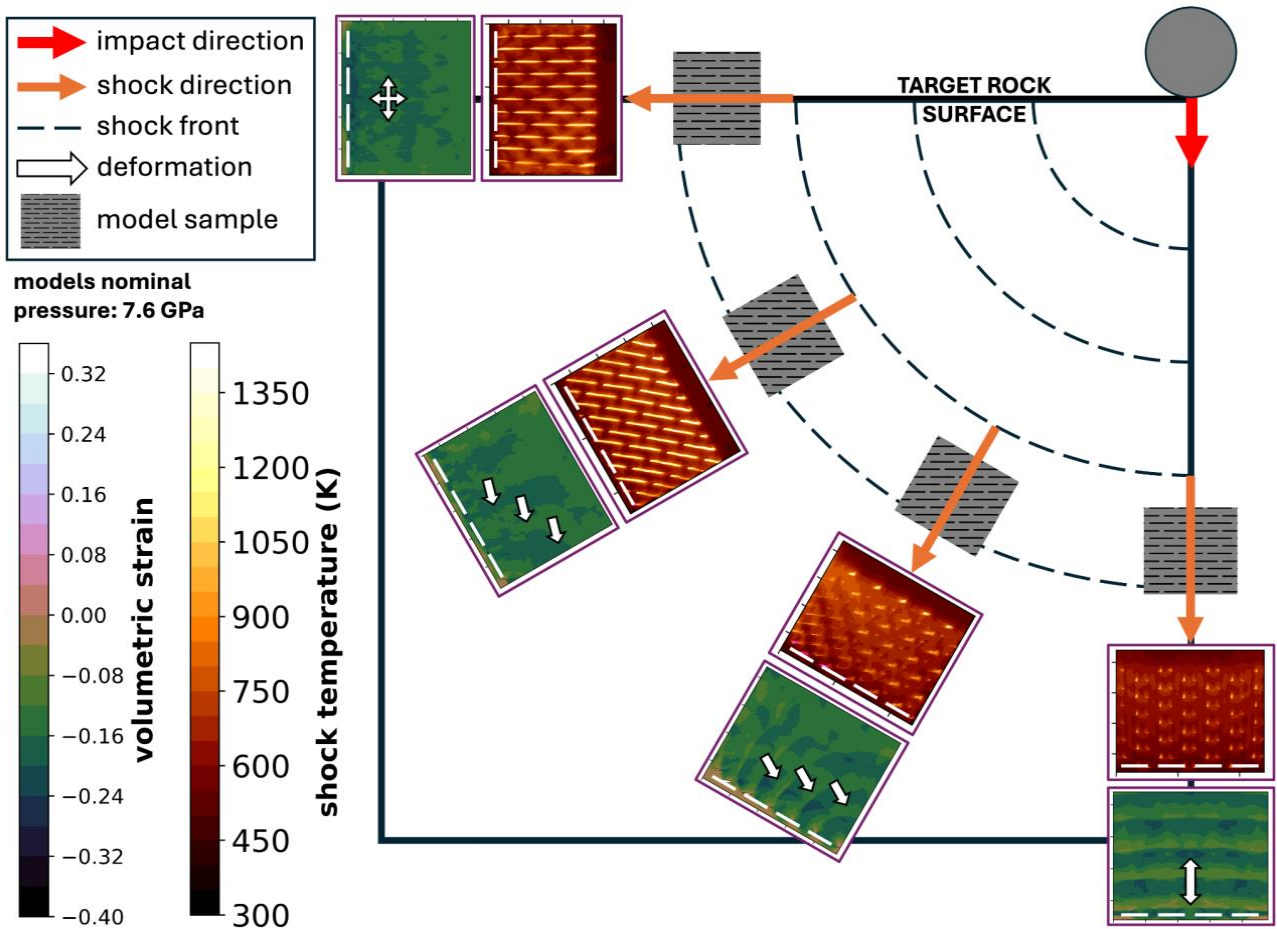

**Fig. 11.** Schematic of an impact on a target rock with added corresponding fenestral porosity models (lineations in carbonate rocks, Fig. 1e) with shock temperature and volumetric strain (deformation) 2-D map distributions. The intensity of shock heating is correlated with the orientation of the lineations to the shock wave front (0°, 30°, 60°, 90°). Deformations are especially given for simulation time steps at the peak of the shock when the shock front reaches the bottom of the sample domain (white dashed line). The nominal pressure is the pressure at the flyer plate (pure carbonate rock). Any interpretation at 90° of orientation does not immediately apply to real impacts because the shock front progressing along the surface of the target rock experiences fast decay from the impact point.

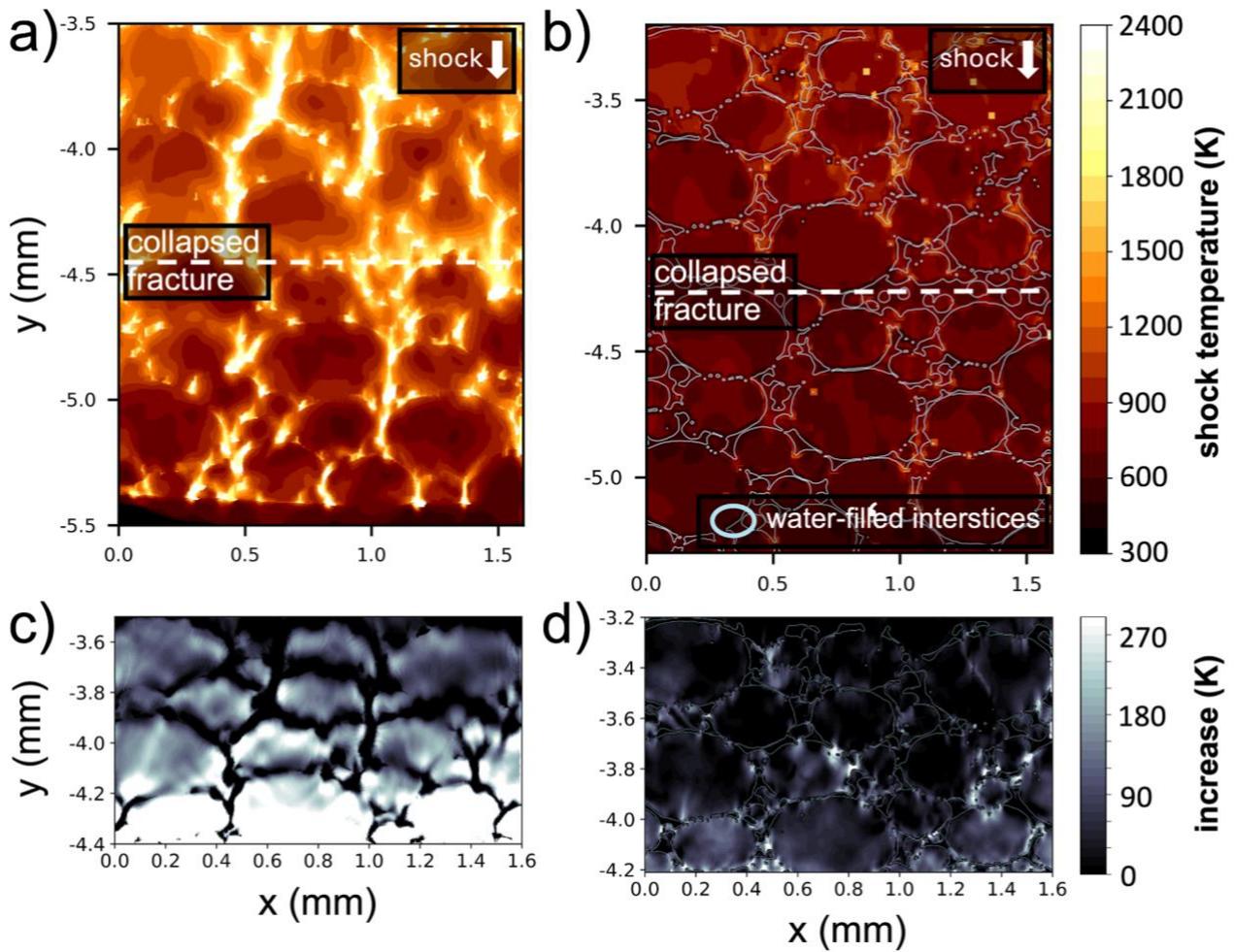

**Fig. 12.** Shock temperature 2-D map distributions within carbonate rock models where a,c) porosity and b,d) water saturation are represented as interstitial space between the carbonate rock particles with bimodal size distribution (Fig. 1f). Higher shock temperatures above the 4.5 mm mark resulted from the collapse of a 120 μm fracture. The increase in shock temperatures is more intense in the vicinity of the collapsed fracture (c). Water saturation substantially reduces the temperature increase from fracture collapse (d). The initial shock front direction is indicated, and models are shown right when the shock wave traverses the sample plate.

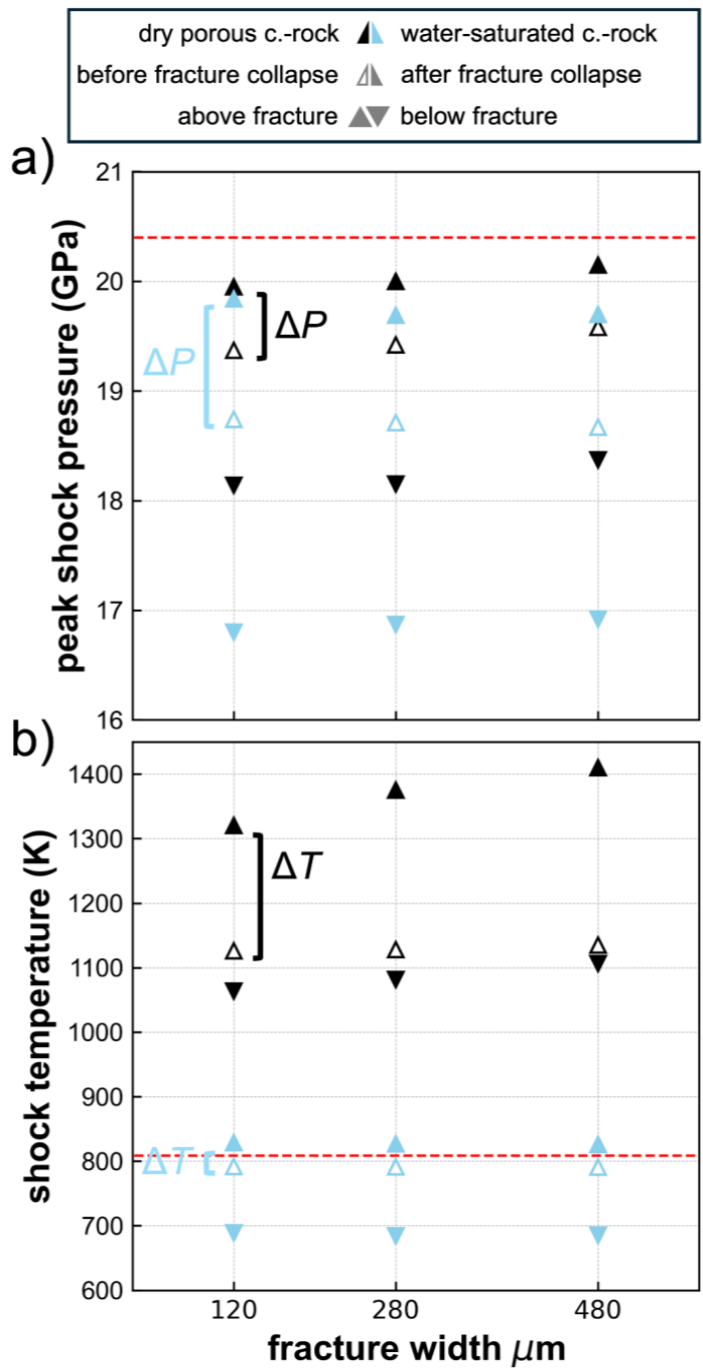

**Fig. 13.** Resulting carbonate rock/water a) bulk peak shock pressures and b) bulk shock temperatures above (point-up triangle) or below (point-down triangle) the fracture as a function of fracture thickness (model shown in Fig. 1f). Temperatures before the collapse of the fracture are represented by empty triangles. Dry carbonate rock and water-saturated carbonate rock are represented by black and blue triangles, respectively. The red dashed lines represent the nominal pressure and shock temperature in the flyer plate in nonporous carbonate rock. The differences in shock pressures ($\Delta P$) and shock temperatures ($\Delta T$) before and after the collapse of the fracture are highlighted for the 120-micron fracture model (Fig. 12). *c.-rock*: carbonate rock.

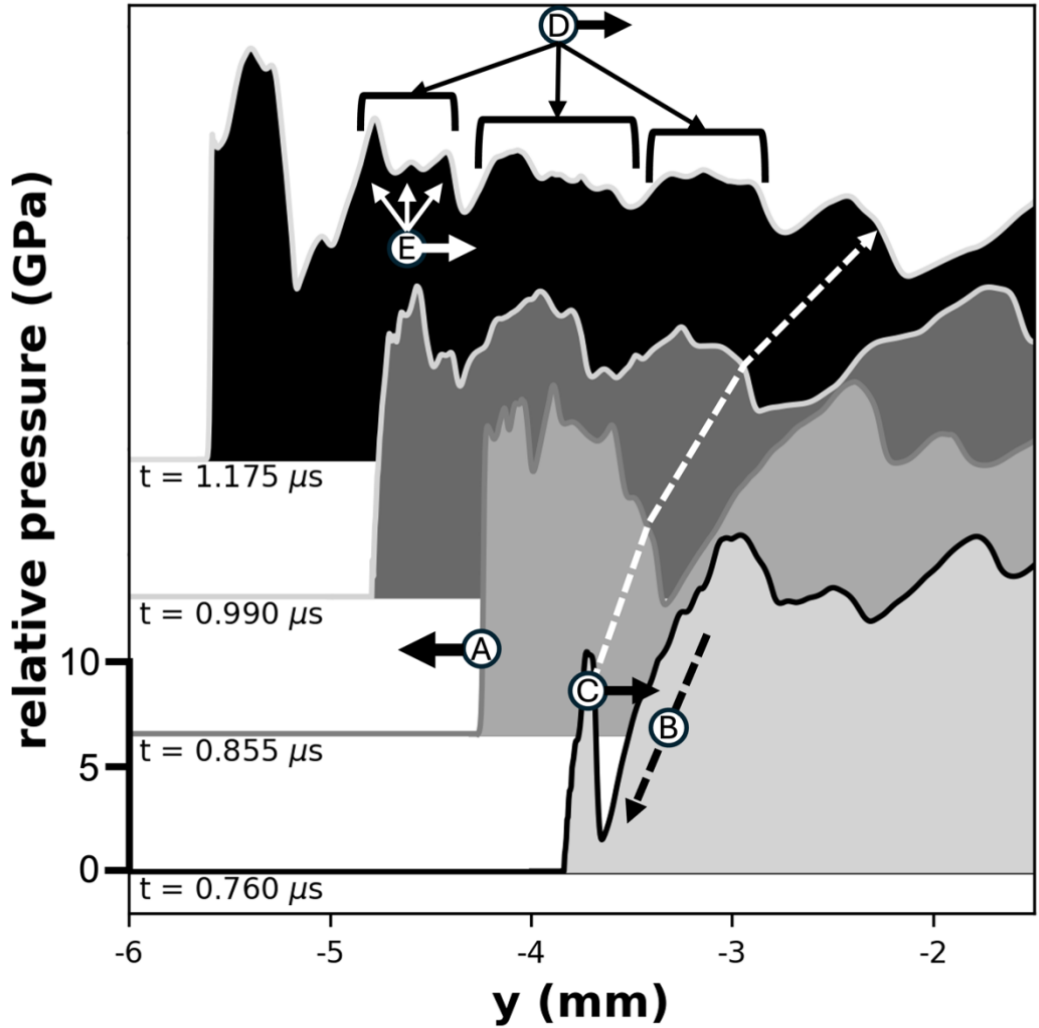

**Fig. 14.** Timewise longitudinal profiles of the pressure field in the 120-micron fracture carbonate rock model with interstitial porosity (Figs. 1f and 12). Patterns observed within the profiles highlight A) the sharp increase of pressure at the shock front, B) the sudden decrease of pressures resulting from the collapse of the fracture with C) the reverberated shock at the complete collapse of the fracture. Interstitial porosity also results in smaller reverberation events up to 3rd order reverberations. Reverberations, in a direction opposite to the shock front, can be traced within each profile, such as depicted by the dashed and white arrowed curve for C), the 1st order reverberation. Each profile is offset by 6.5 GPa for better visualization.

## 1. On the use of a cylindrical axis of symmetry

Cylindrical symmetry is used to simulate impact models where the projectile hits at 90° head-on the target rock. Using a cylindrical axis of symmetry in models enables the study of impact cratering without the need for 3D modeling, especially mandatory for oblique impacts (Littlefield and Dawson, 2006). When calculations are cylindrically projected, the shock wave and resulting pressures both follow suit by adopting similar cylindrical and radial propagation into the target. In mesoscale models, using a planar setup will lead to an underestimation of shock pressures in a single particle placed on the axis (Fig. 1c; Moreau et al. 2018), and a cylindrical axis of symmetry is, thus, recommended.

However, biases are introduced in mesoscale models if an axis of cylindrical symmetry is applied to layers that contain several particles situated in different areas of the sample layer (e.g., Fig. 1b). In cylindrical models, particles that are not exactly centered on the axis of symmetry adopt a torus shape, which radius depends on the distance from the axis of symmetry (Güldemeister et al., 2013). This means that particles situated near the axis of symmetry affect the shock wave as torus-like material, and particles far from the axis of symmetry affect the shock wave as rod-like material. The consequence is that models adopt a skewed distribution of pressures. Complementary tests made on the mesoscale models of this work show that this skewed distribution of pressures leads to a reasonable balancing of results compared to exact planar replicas of the models (Fig. S5–S7). Shock wave and particle velocity fields are also shown in the figures, but they are only of informative value for comparing models.

Littlefield D. L., and Dawson A. 2006. The role of impactor shape and obliquity on crater evolution in celestial impacts. *Int. J. Impact Eng.* **33**, 371–379. https://doi.org/10.1016/j.ijimpeng.2006.09.037.

**Table S1. Dolostone iSALE material parameters.**

| | |
|---|---|
| **porosity model** | ε–α compaction |
| distention ($1/1\text{-}\Phi_{porosity}$) | 1 |
| | |
| **strength model** | Collins et al. (2004) |
| Poisson's ratio | 0.3 |
| | |
| *intact* | |
| cohesive strength (*yint0*, Pa) | $9 \times 10^6$ |
| coefficient of friction (*fricint*) | 0.65 |
| limited strength (*ylimint*, Pa) | $2.6 \times 10^9$ |
| | |
| *damaged* | |
| cohesive strength (*ydam0*, Pa) | $0.09 \times 10^6$ |
| coefficient of friction (*fricdam*) | 0.44 |
| limited strength (*ylimdam*, Pa) | $2.6 \times 10^9$ |
| | |
| **damage model** | |
| minimum failure strain (*ivanov_a*) | $1 \times 10^{-4}$ |
| constant (*ivanov_b*) | $1 \times 10^{-11}$ |
| compressional failure (*ivanov_c*, *Pa*) | $3 \times 10^{+8}$ |
| | |
| **thermal properties** | Ohnaka (1995) |
| thermal softening (tfrac) | 1.2 |
| melting temperature (K) | 1500 |
| heat capacity (J/kgK) | 892 |

\* additional parameters for strength/porosity/damage models in the *Research Data*.

**Table S2. Mesoscale models, general results (high pressure).**

| % *water* | % pores | nominal pressure (GPa)[a] | peak shock pressure (GPa) | | peak shock temperature (K) | | density (kg/m³) | |
|---|---|---|---|---|---|---|---|---|
| **Model parameters** | | **Results** | | | | | | |
| *10% water (Tillotson EoS)* | | | | | | | | |
| 9.73 | - | 4.45 | 6.41 | (0.82) | 357 | (41) | 2881 | (386) |
| 9.99 | - | 7.56 | 10.61 | (1.03) | 433 | (66) | 3031 | (368) |
| 10.27 | - | 11.25 | 15.78 | (2.17) | 512 | (80) | 3167 | (371) |
| 9.98 | - | 15.54 | 21.10 | (2.62) | 611 | (95) | 3287 | (365) |
| 10.14 | - | 20.40 | 26.88 | (2.88) | 745 | (100) | 3437 | (372) |
| 9.89 | - | 32.27 | 40.11 | (4.68) | 1058 | (150) | 3707 | (370) |
| 10.1 | - | 45.80 | 55.53 | (5.43) | 1497 | (177) | 3894 | (386) |
| 10.29 | - | 60.22 | 73.19 | (6.88) | 2046 | (201) | 4140 | (400) |
| 10.1 | - | 77.52 | 97.50 | (13.37) | 2835 | (278) | 4304 | (420) |
| 9.85 | - | 95.76 | 129.93 | (39.01) | 3872 | (1046) | 4544 | (448) |
| | | | | | | | | |
| *10% water (ANEOS)* | | | | | | | | |
| 9.85 | - | 4.45 | 5.66 | (1.09) | 416 | (40) | 2838 | (341) |
| 10.16 | - | 7.56 | 9.57 | (1.51) | 503 | (56) | 2992 | (348) |
| 10.14 | - | 11.25 | 14.15 | (2.20) | 601 | (79) | 3141 | (367) |

**Table S2. Mesoscale models, general results (high pressure).**

| % water | % pores | nominal pressure (GPa)[a] | peak shock pressure (GPa) | | peak shock temperature (K) | | density (kg/m³) | |
|---|---|---|---|---|---|---|---|---|
| **Model parameters** | | | **Results** | | | | | |
| *10% water (Tillotson EoS)* | | | | | | | | |
| 10.09 | - | 15.54 | 19.17 | (3.21) | 733 | (106) | 3264 | (373) |
| 9.67 | - | 20.40 | 24.61 | (3.45) | 878 | (105) | 3407 | (376) |
| 10.03 | - | 32.27 | 37.06 | (4.11) | 1251 | (132) | 3642 | (402) |
| 10.12 | - | 45.80 | 52.80 | (7.34) | 1764 | (177) | 3856 | (418) |
| 9.93 | - | 60.22 | 71.42 | (7.90) | 2419 | (233) | 4031 | (454) |
| 9.87 | - | 77.52 | 99.77 | (28.61) | 3271 | (679) | 4233 | (453) |
| 10.02 | - | 95.76 | 122.78 | (30.51) | 4095 | (797) | 4459 | (459) |
| *20% water (ANEOS)* | | | | | | | | |
| 19.55 | - | 4.45 | 5.52 | (0.97) | 428 | (48) | 2709 | (465) |
| 19.83 | - | 7.56 | 10.10 | (1.78) | 524 | (77) | 2893 | (475) |
| 20.08 | - | 11.25 | 14.39 | (2.60) | 635 | (108) | 3000 | (487) |
| 19.61 | - | 15.54 | 19.93 | (2.64) | 765 | (126) | 3151 | (503) |
| 19.63 | - | 20.40 | 25.56 | (4.32) | 936 | (147) | 3279 | (518) |
| 20.1 | - | 32.27 | 38.92 | (5.55) | 1353 | (191) | 3490 | (548) |
| 19.89 | - | 45.80 | 55.60 | (10.48) | 1878 | (308) | 3706 | (567) |
| 20.03 | - | 60.22 | 78.00 | (16.56) | 2548 | (364) | 3838 | (594) |
| 19.82 | - | 77.52 | 104.44 | (37.83) | 3355 | (999) | 4058 | (619) |
| 19.67 | - | 95.76 | 130.80 | (43.41) | 4178 | (1383) | 4214 | (626) |
| *40% water (ANEOS)* | | | | | | | | |
| 39.28 | - | 4.45 | 5.63 | (1.05) | 443 | (54) | 2479 | (613) |
| 38.62 | - | 7.56 | 9.85 | (1.94) | 544 | (92) | 2630 | (615) |
| 39.27 | - | 11.25 | 14.52 | (2.49) | 661 | (133) | 2753 | (640) |
| 39.44 | - | 15.54 | 19.55 | (3.84) | 803 | (163) | 2856 | (655) |
| 38.73 | - | 20.40 | 25.50 | (5.33) | 987 | (215) | 2968 | (663) |
| 39.08 | - | 32.27 | 39.56 | (8.82) | 1403 | (287) | 3169 | (695) |
| 38.89 | - | 45.80 | 55.37 | (8.69) | 1915 | (309) | 3359 | (728) |
| 39.02 | - | 60.22 | 79.21 | (27.21) | 2616 | (865) | 3486 | (747) |
| 38.94 | - | 77.52 | 100.98 | (32.46) | 3315 | (1133) | 3651 | (765) |
| 39.19 | - | 95.76 | 128.05 | (38.24) | 4112 | (1665) | 3836 | (809) |

\* values for individual phases (carbonate rock/water) of specific models are compiled in the Research Data. Values in parentheses are standard variations.
(a) pressure recorded in the flyer layer of the pure carbonate rock model.
(b) values are recorded when the shock wave occupies the entirety of the sample layer at chosen time step.
(c) approximately calculated as follows at chosen time step: (distance from collision site) / (time step duration).

**Table S2. Mesoscale models, general results (high pressure) (continued).**

| % water | % pores | nominal pressure (GPa)[a] | peak shock pressure (GPa) | | peak shock temperature (K) | | density (kg/m³) | |
|---|---|---|---|---|---|---|---|---|
| **Model parameters** | | | **Results** | | | | | |
| *33% water (ANEOS)* | | | | | | | | |
| 33.16 | - | 4.45 | 4.99 | (1.49) | 579 | (146) | 2621 | (75) |
| 33.16 | - | 11.25 | 15.60 | (5.87) | 1152 | (417) | 2981 | (104) |
| 33.16 | - | 32.27 | 48.52 | (20.75) | 2656 | (1670) | 3370 | (247) |
| 33.16 | - | 60.22 | 89.50 | (32.70) | 4255 | (2796) | 3751 | (342) |
| 33.16 | - | 95.76 | 138.85 | (51.05) | 6086 | (3633) | 4066 | (356) |
| *18% water, 15% pores (ANEOS)* | | | | | | | | |
| 18.09 | 15.07 | 4.45 | 5.67 | (1.76) | 510 | (127) | 2564 | (475) |
| 18.09 | 15.07 | 7.56 | 10.18 | (3.60) | 680 | (248) | 2764 | (467) |
| 18.09 | 15.07 | 11.25 | 15.39 | (5.20) | 865 | (347) | 2888 | (475) |
| 18.09 | 15.07 | 20.40 | 28.56 | (11.21) | 1324 | (644) | 3115 | (512) |
| 18.09 | 15.07 | 32.27 | 45.80 | (20.31) | 1907 | (1048) | 3314 | (546) |
| 18.09 | 15.07 | 45.80 | 65.56 | (28.81) | 2586 | (1686) | 3474 | (579) |
| 18.09 | 15.07 | 60.22 | 88.10 | (37.27) | 3308 | (2445) | 3647 | (619) |
| 18.09 | 15.07 | 77.52 | 112.45 | (45.38) | 4033 | (2569) | 3842 | (632) |
| 18.09 | 15.07 | 95.76 | 138.04 | (54.17) | 4847 | (2895) | 3982 | (675) |
| *17% pores (ANEOS)* | | | | | | | | |
| 0 | 16.95 | 4.45 | 5.92 | (1.47) | 528 | (116) | 2783 | (69) |
| 0 | 16.95 | 11.25 | 15.94 | (5.04) | 898 | (293) | 3113 | (76) |
| 0 | 16.95 | 32.27 | 50.07 | (24.17) | 2017 | (976) | 3574 | (138) |
| 0 | 16.95 | 60.22 | 97.94 | (43.13) | 3545 | (1819) | 3956 | (290) |
| 0 | 16.95 | 95.76 | 152.99 | (56.79) | 5162 | (2739) | 4331 | (323) |
| *100% carbonate rock (calcite, ANEOS)* | | | | | | | | |
| | | (in *Research Data*) | | | | | | |
| *100% water (ANEOS)* | | | | | | | | |
| | | (in *Research Data*) | | | | | | |

\* values for individual phases (carbonate rock/water) of specific models are compiled in the Research Data.
(a) pressure recorded in the flyer layer of the pure carbonate rock model.
(b) values are recorded when the shock wave occupies the entirety of the sample layer at chosen time step.
(c) approximately calculated as follows at chosen time step: (distance from collision site) / (time step duration).

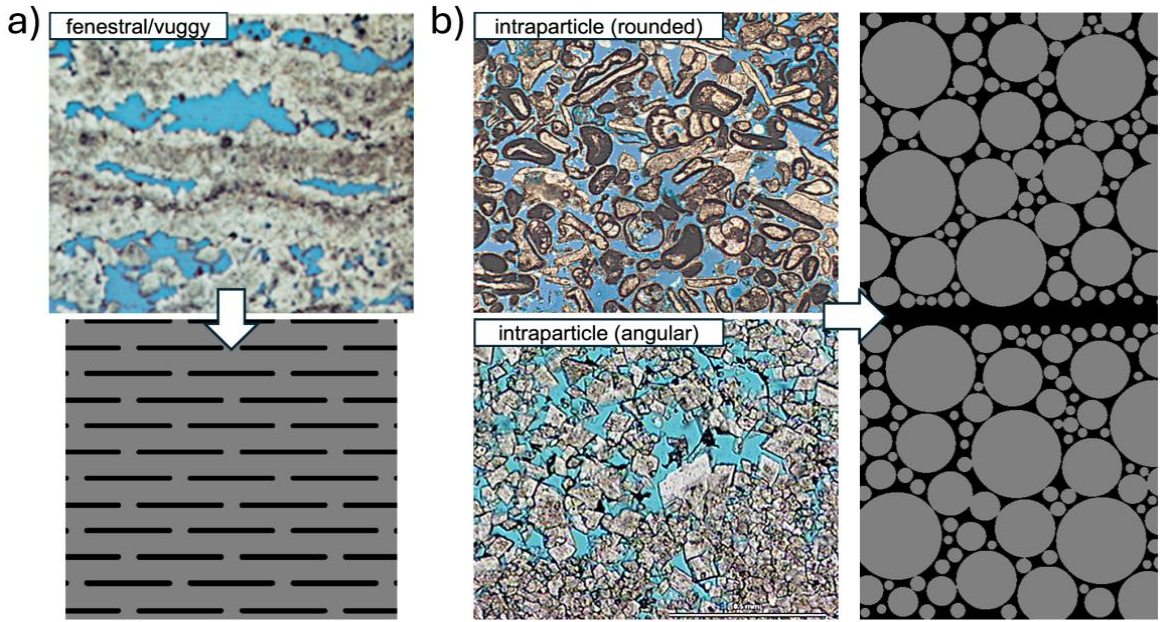

**Fig. S1.** Examples of real porosity textures found in carbonate rocks and the simplified associated 2D iSALE models: a) fenestral/vuggy porosity (Plummer 2021, mod.) translated to the lineation models, b) interparticle porosity (Carbonateworld©, mod.) translated to the open fracture interstitial porosity model. Blue areas in thin section pictures are highlighted pore space.

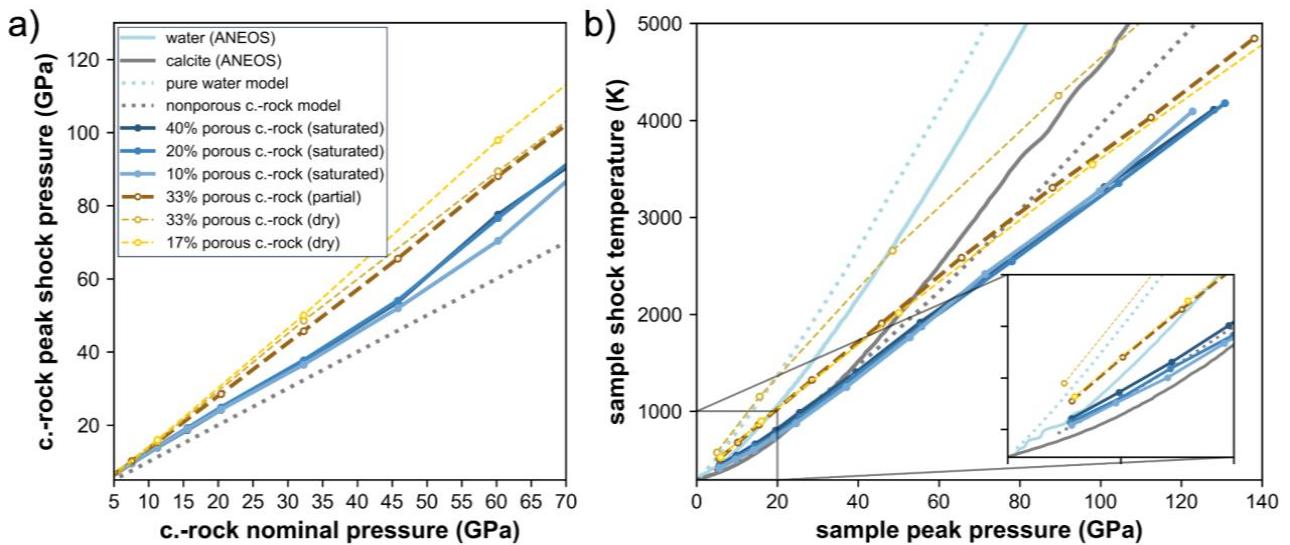

**Fig. S2**. Influence of water saturation on a) carbonate rock peak shock pressure and b) sample shock temperature in function of the carbonate rock nominal pressure and sample peak shock pressure, respectively. ANEOS is used for any plots shown. The carbonate rock is based on ANEOS calcite. Nominal pressure: pressure recorded in the carbonate rock flyer plate in a model devoid of water/pore particles. When comparing to Fig. 3, *Main Document*, we see that a normalization by nominal pressures is necessary to understand exactly how water or pores influence temperatures; here, panel b) shows that temperatures are reduced when water fills the pores, but in reality, the pressure regime (a) is ultimately different and thus a nominal pressure is required for comparing models.

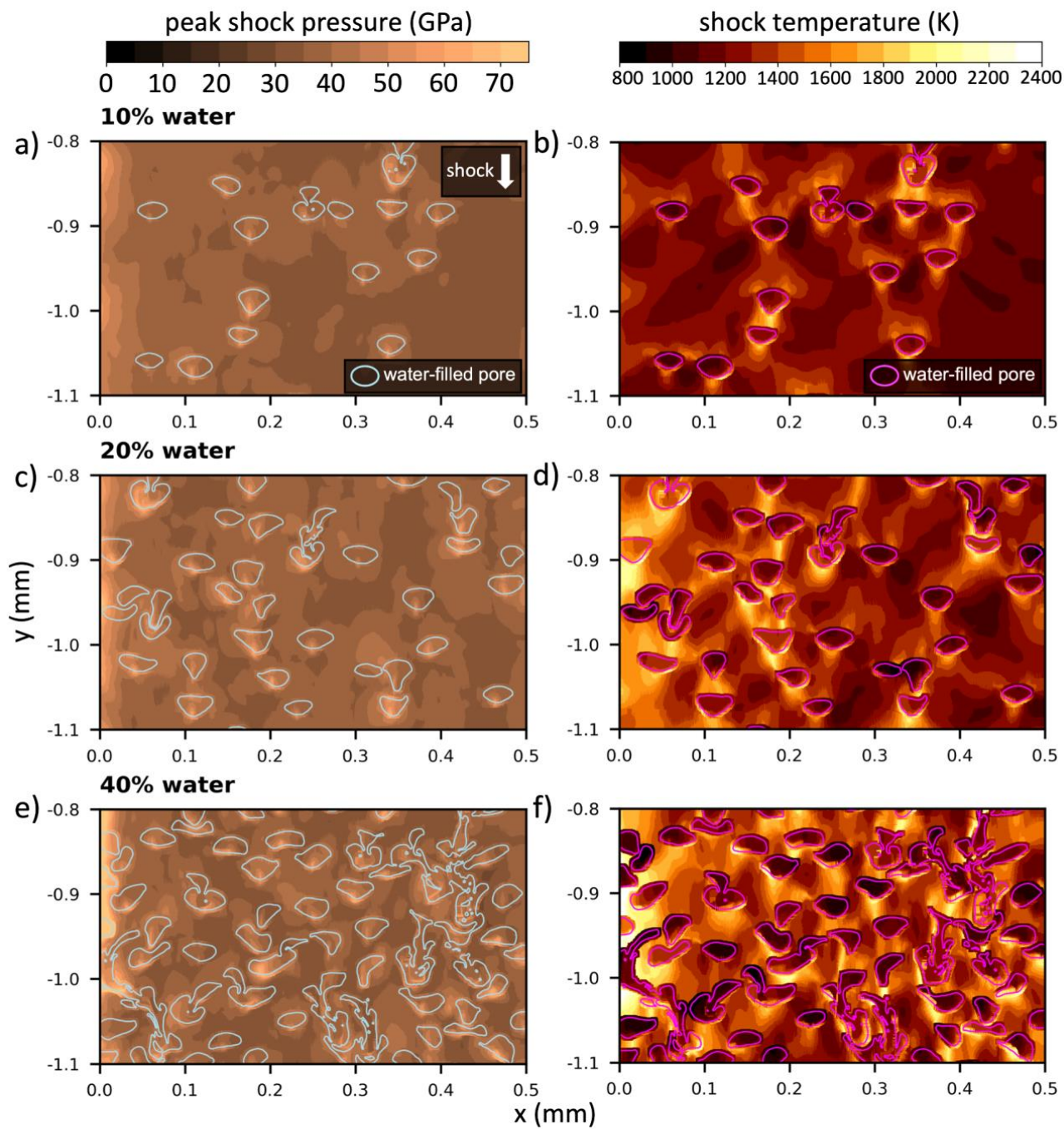

**Fig. S3.** Water saturation effect on the distribution of peak shock pressures and shock temperatures in the carbonate rock at high shock pressure (~32 GPa nominal pressure). Legend indicates positions of water particles (20 cells per particle radius). The initial shock front direction is indicated, and models are shown right after passage of the shock wave.

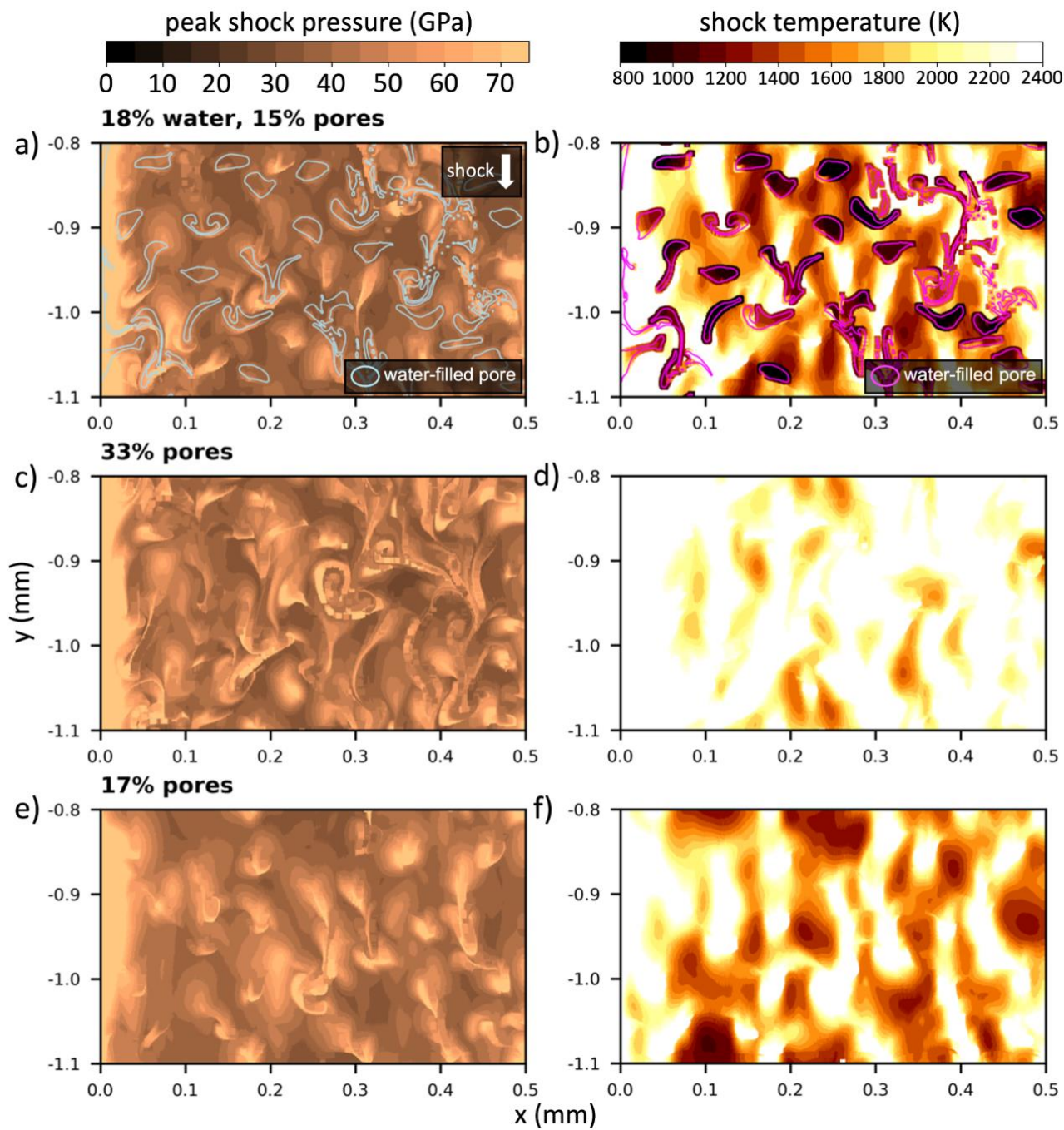

**Fig. S4.** Water saturation effect on the distribution of peak shock pressures and shock temperatures in the carbonate rock at high shock pressure (~32 GPa nominal pressure) Legend indicates positions of water particles (20 cells per particle radius). Pores are not apparent because they are closed from shock. Pores were filled with void rather than air as usually seen in mesoscale models (sandstone, Güldemeister et al., 2013). The initial shock front direction is indicated, and models are shown right after passage of the shock wave.

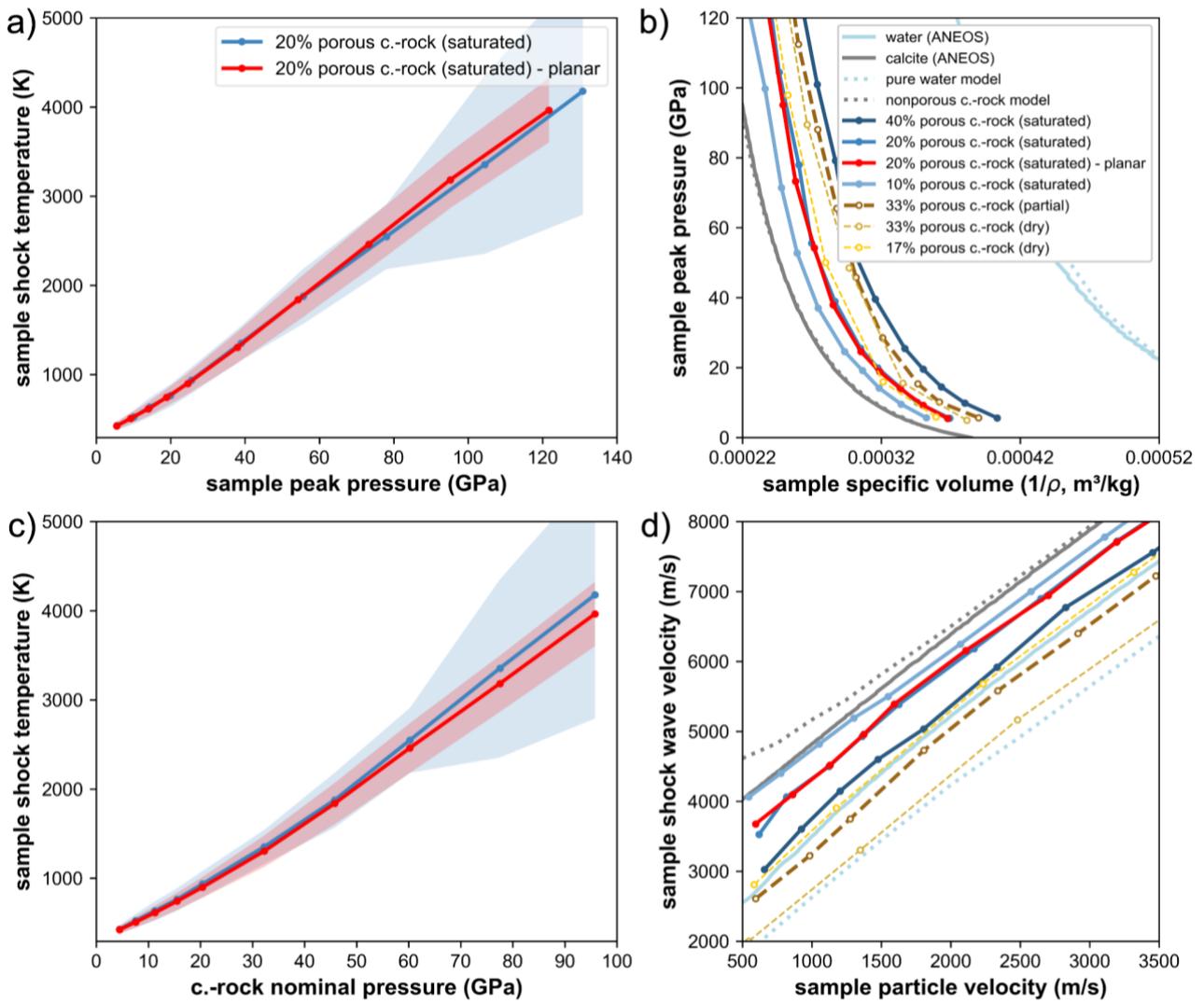

**Fig. S5.** Comparison between using a cylindrical axis of symmetry and planar model for the 20% water-saturated carbonate rock model, where water is resolved as non-connected water-filled pores. Although a,c) distribution of temperatures flares up with increasing pressures (see standard deviation shaded areas), the general increase of temperatures and shape of the curves are similar in either models. Compiled Hugoniot data (b,d) do not show any discrepancies between either use of a cylindrical axis or planar model. In cylindrical symmetry, shock wave interactions are more intense at the axis because of pores adopting a toroidal shape, hence the stronger variation of temperatures observed in a,c).

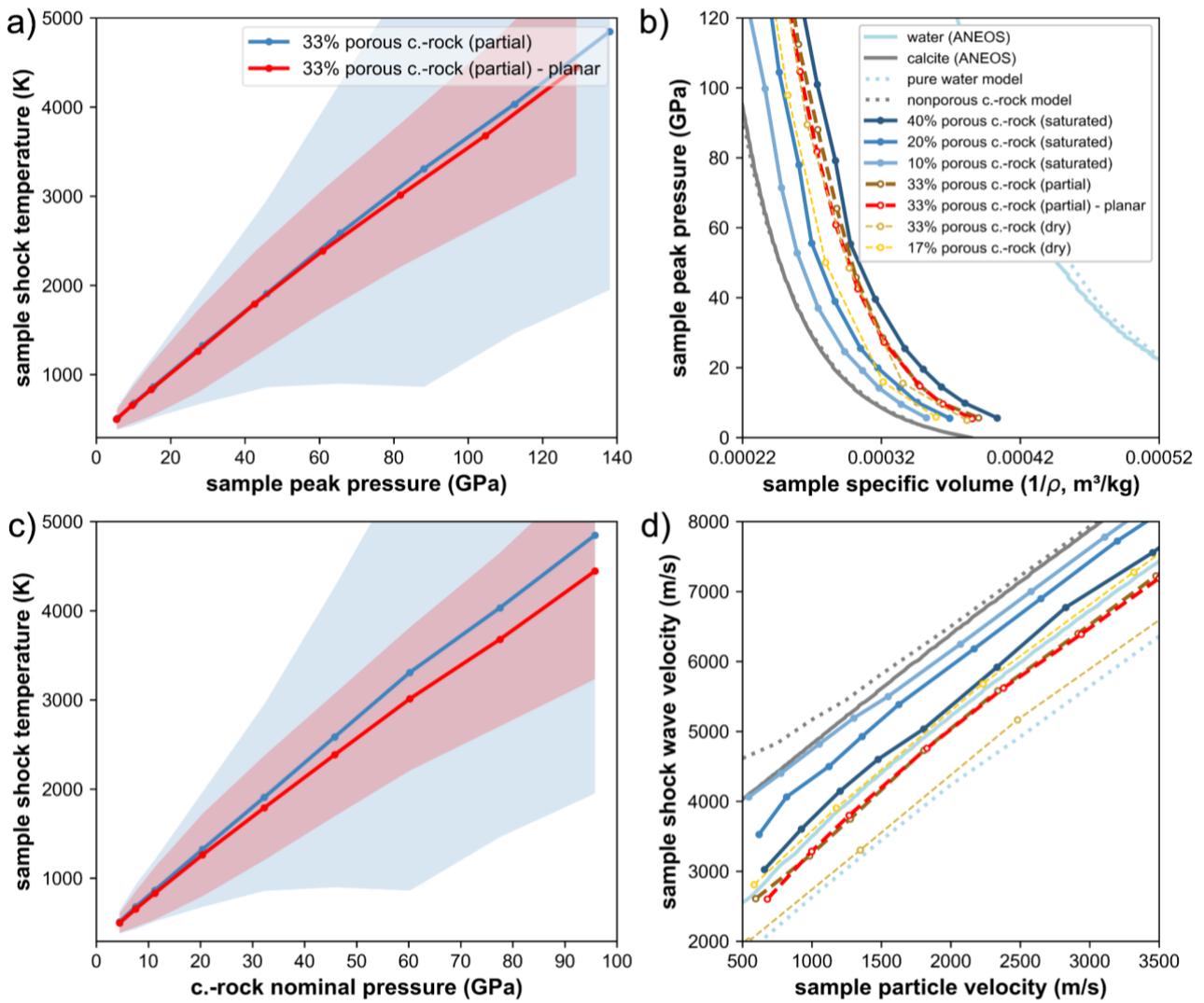

**Fig. S6.** Comparison between using a cylindrical axis of symmetry and planar model for the partially water-saturated carbonate rock model, where water is resolved as non-connected water-filled pores, as are empty pores. Although a,c) distribution of temperatures flares up with increasing pressures (see standard deviation shaded areas), the general increase of temperatures and shape of the curves are similar in either models. Compiled Hugoniot data (b,d) do not show any discrepancies between either use of a cylindrical axis or planar model. In cylindrical symmetry, shock wave interactions are more intense at the axis because of pores adopting a toroidal shape, hence the stronger variation of temperatures observed in a,c).

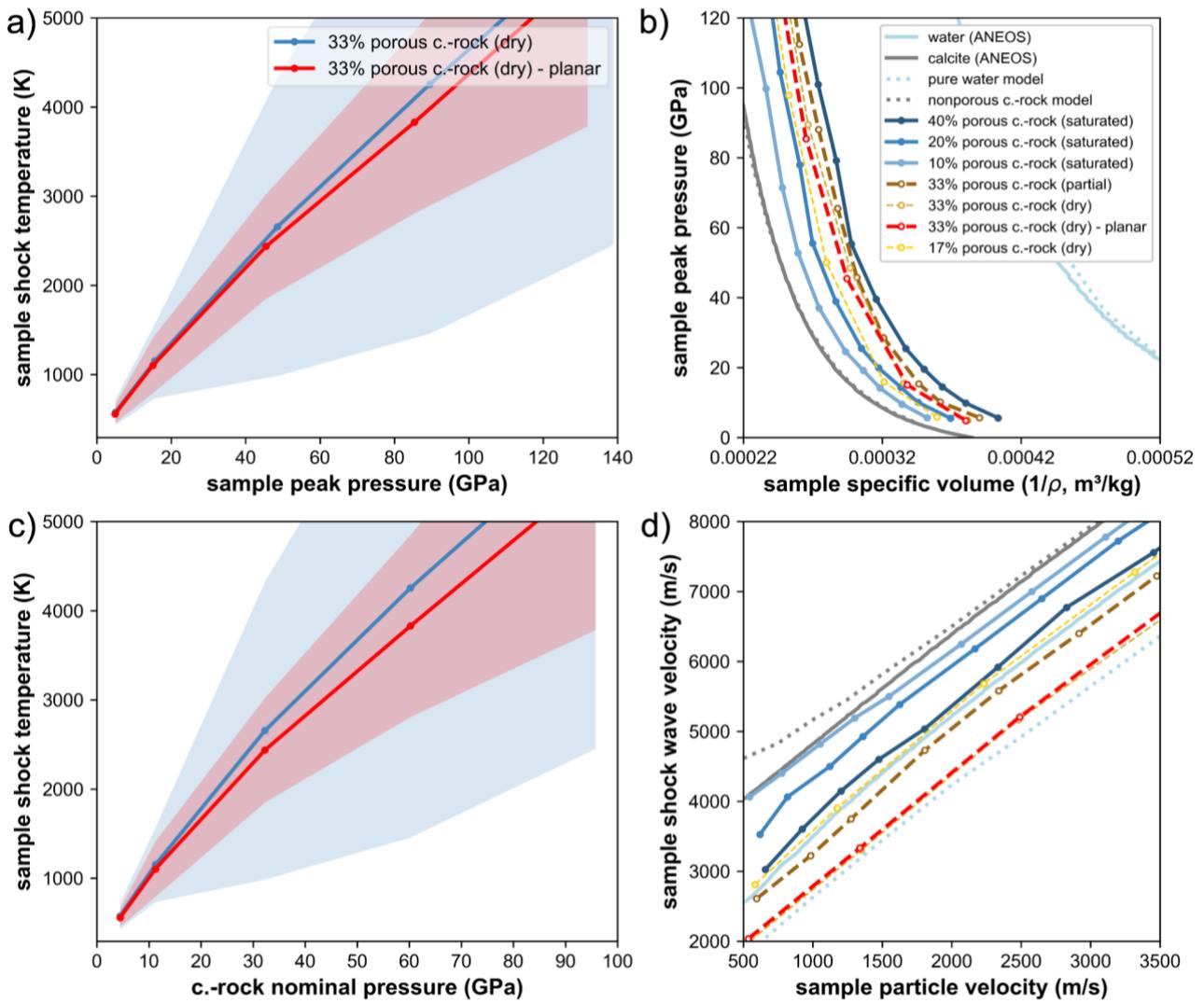

**Fig. S7.** Comparison between using a cylindrical axis of symmetry and planar model for the dry porous carbonate rock model, where porosity is resolved as non-connected pores. Although a,c) distribution of temperatures flares up with increasing pressures (see standard deviation shaded areas), the general increase of temperatures and shape of the curves are similar in either models. Compiled Hugoniot data (b,d) do not show any discrepancies between either use of a cylindrical axis or planar model. In cylindrical symmetry, shock wave interactions are more intense at the axis because of pores adopting a toroidal shape, hence the stronger variation of temperatures observed in a,c).